\documentclass[sigconf, nonacm]{acmart}

\usepackage[normalem]{ulem}
\usepackage{amsmath}
\usepackage{algorithm}
\usepackage[noend]{algpseudocode}
\usepackage{subfig}
\usepackage{enumitem}
\newcommand{\parahead}[1]{\textbf{\textit{#1}}}
\def \SYS{Flash-LLM} 
\usepackage{titlesec}

\titlespacing*% the star= don't indent first paragraph after
    {\subsection}% which command you want to set the spacing for
    {0pt}% spacing to the left of heading
    {1ex}% spacing before the heading
    {1ex}% spacing after the heading
\titlespacing*%
    {\section}% 
    {0pt}% 
    {1ex}% 
    {1ex}%
\titlespacing*%
    {\subsubsection}% 
    {0pt}% 
    {1ex}% 
    {1ex}%
\usepackage{marginnote}
\newcommand\vldbdoi{XX.XX/XXX.XX}

\newcommand\vldbavailabilityurl{}
\begin{document}

%%%% reduce the space around equations
\setlength{\abovedisplayskip}{2pt}
\setlength{\belowdisplayskip}{2pt}
%%%%
%%%%%%%% figure/table float
\setlength{\floatsep}{5pt}
\setlength{\textfloatsep}{5pt}
\setlength{\intextsep}{3pt}
%%%%%%%%

\title{
\SYS: Enabling Cost-Effective and Highly-Efficient Large Generative Model Inference with Unstructured Sparsity 
% \SYS: Enabling Low-cost and Ultra-efficient Large Generative Model Inference through Tensor Core Support for Unstructured Sparsity
}

%%
%% The "author" command and its associated commands are used to define the authors and their affiliations.

\author{Haojun Xia $^* \dagger$} 
\affiliation{%
  \institution{Univeristy of Sydney}
  % \city{Sydney}
  % \state{Australia}
}
\email{hxia6845@uni.sydney.edu.au}

\author{Zhen Zheng $^*$}
\affiliation{%
  \institution{Alibaba Group}
  % \city{Hangzhou}
  % \state{China}
}
\email{james.zz@alibaba-inc.com}

\author{Yuchao Li}
\affiliation{%
  \institution{Alibaba Group}
  % \city{Hangzhou}
  % \state{China}
}
\email{laiyin.lyc@alibaba-inc.com}

\author{Donglin Zhuang}
\affiliation{%
  \institution{Univeristy of Sydney}
  % \city{Sydney}
  % \state{Australia}
}
\email{donglin.zhuang@sydney.edu.au}

\author{Zhongzhu Zhou }
\affiliation{%
  \institution{Univeristy of Sydney}
  % \city{Sydney}
  % \state{Australia}
}
\email{zhongzhu.zhou@sydney.edu.au}

\author{Xiafei Qiu}
\affiliation{%
  \institution{Alibaba Group}
  % \city{Hangzhou}
  % \state{China}
}
\email{xiafei.qiuxf@alibaba-inc.com}

\author{Yong Li}
\affiliation{%
  \institution{Alibaba Group}
  % \city{Hangzhou}
  % \state{China}
}
\email{jiufeng.ly@alibaba-inc.com}

\author{Wei Lin}
\affiliation{%
  \institution{Alibaba Group}
  % \city{Hangzhou}
  % \state{China}
}
\email{weilin.lw@alibaba-inc.com}

\author{Shuaiwen Leon Song}
\affiliation{%
  \institution{Univeristy of Sydney}
  % \city{Sydney}
  % \state{Australia}
}
\email{shuaiwen.song@sydney.edu.au}

%%
%% The abstract is a short summary of the work to be presented in the
%% article.
\begin{abstract}

With the fast growth of parameter size, it becomes increasingly challenging to deploy large generative models as they typically require large GPU memory consumption and massive computation.
Unstructured model pruning has been a common approach to reduce both GPU memory footprint and the overall computation while retaining good model accuracy.
However, the existing solutions do not provide a highly-efficient support for handling unstructured sparsity on modern GPUs, especially on the highly-structured tensor core hardware.
Therefore, we propose \SYS{} for enabling low-cost and highly-efficient large generative model inference with the sophisticated support of unstructured sparsity on high performance but highly restrictive tensor cores.
Based on our key observation that the main bottleneck of generative model inference is the several skinny matrix multiplications for which tensor cores would be significantly under-utilized due to low computational intensity,
we propose a general \textit{Load-as-Sparse and Compute-as-Dense} methodology for unstructured sparse matrix multiplication (SpMM).
The basic insight is to address the significant memory bandwidth bottleneck while tolerating redundant computations that are not critical for end-to-end performance on tensor cores.
%Based on this, we design an effective software framework for tensor core based unstructured SpMM, leveraging on-chip resources for efficient sparse data extraction and coordination, fast dense feature-map data loading, and multi-level computation overlapping and pipelining. 
Based on this, we design an effective software framework for tensor core based unstructured SpMM, leveraging on-chip resources for efficient sparse data extraction and computation/memory-access overlapping. 
Extensive evaluations demonstrate that (1) at SpMM kernel level, \SYS{} significantly outperforms the state-of-the-art library, i.e., Sputnik and SparTA by an average of $2.9 \times$ and $1.5 \times$, respectively.(2) At end-to-end framework level on OPT-30B/66B/175B models, for \textit{tokens per GPU-second}, \SYS{} achieves up to $3.8 \times$ and $3.6 \times$ improvement over DeepSpeed and FasterTransformer, respectively, with significantly lower inference cost.
\SYS’s source code is publicly available at \url{https://github.com/AlibabaResearch/flash-llm}.

\end{abstract}

\maketitle

% %%% do not modify the following VLDB block %%
% %%% VLDB block start %%%
% \pagestyle{\vldbpagestyle}
% \begingroup\small\noindent\raggedright\textbf{PVLDB Reference Format:}\\
% \vldbauthors. \vldbtitle. PVLDB, \vldbvolume(\vldbissue): \vldbpages, \vldbyear.\\
% \href{https://doi.org/\vldbdoi}{doi:\vldbdoi}
% \endgroup
% \begingroup
% \renewcommand\thefootnote{}\footnote{\noindent
% This work is licensed under the Creative Commons BY-NC-ND 4.0 International License. Visit \url{https://creativecommons.org/licenses/by-nc-nd/4.0/} to view a copy of this license. For any use beyond those covered by this license, obtain permission by emailing \href{mailto:info@vldb.org}{info@vldb.org}. Copyright is held by the owner/author(s). Publication rights licensed to the VLDB Endowment. \\
% \raggedright Proceedings of the VLDB Endowment, Vol. \vldbvolume, No. \vldbissue\ %
% ISSN 2150-8097. \\
% \href{https://doi.org/\vldbdoi}{doi:\vldbdoi} \\
% }\addtocounter{footnote}{-1}\endgroup
% %%% VLDB block end %%%

% %%% do not modify the following VLDB block %%
% %%% VLDB block start %%%
% \ifdefempty{\vldbavailabilityurl}{}{
% \vspace{.3cm}
% \begingroup\small\noindent\raggedright\textbf{PVLDB Artifact Availability:}\\
% The source code, data, and/or other artifacts have been made available at \url{\vldbavailabilityurl}.
% \endgroup
% }
% %%% VLDB block end %%%

%%% contribution and intern information %%%
{\let\thefootnote\relax\footnote{{
$^*$ Haojun and Zhen contributed equally.
}}}
{\let\thefootnote\relax\footnote{{
$\dagger$ Part of the work was done when Haojun was interned at Alibaba.
}}}
\setcounter{footnote}{0} % reset footnote number
%%% contribution and intern information end %%%

%%% Main Chapters start %%%
\vspace{-6ex} % To delete in the final paper.

\section{Introduction}
\label{sec:introduction}

%In recent years, the parameter size of generative models is growing rapidly (GPT-2~\cite{GPT-2} 1.5 billion parameters, GPT-3~\cite{GPT-3} 175 billion, and Megatron-Turing NLG~\cite{Megatron-Turing-NLG} 530 billion).
Generative models have demonstrated their effectiveness across a wide range of language and data management tasks~\cite{GPT-2, GPT-3, 10.14778/3554821.3554896, 10.14778/3574245.3574258, 10.14778/3514061.3514067}. 
However, with the rapid growth of the parameter size (e.g. GPT-2~\cite{GPT-2} 1.5 billion parameters, GPT-3~\cite{GPT-3} 175 billion, and Megatron-Turing NLG~\cite{Megatron-Turing-NLG} 530 billion),
it becomes increasingly challenging to efficiently deploy these models.
% \cb{
% \marginpar{\cb{R3.D1}}
On one hand, their weights could be too large to be placed on GPUs.
For example, GPT-3 model requires 350GB memory to store only the parameters with FP16 data type, whereas the NVIDIA A100 GPU~\cite{Ampere_WhitePaper} only has a max of 80 GB memory.
On the other hand, large generative models usually cause very high inference latency even using multiple GPUs as large amounts of computation and memory access are required.
% }
%(e.g., tens of seconds for GPT-3 inference with a sequence length of 512).

% There are three basic requirements for large model inference serving: accuracy, latency (and throughput), and cost (i.e., how much hardware resource it consumes).
% The giant model usually consumes too much memory and needs multiple devices to serve for a single model, resulting in how cost and low ROI (return of investment).
% The massive memory accesses and computations also lead to high latency.
% Fortunately, the weight pruning approach~\cite{Model_Compression} is demonstrated to be effective to reduce the memory usage and computations for model inference while retaining a good accuracy, by removing a portion of less salient connections in neural networks.
% As a result, the latency increases and the cost decreases.

There are three basic characteristics for practical model inference: accuracy, efficiency (i.e., latency and throughput), and cost (i.e., how much hardware resource it consumes).
The common approach to deploy large models by partitioning the model weights onto multiple devices~\cite{Megatron-LM, zheng2022alpa} could suffer from high cost and low efficiency.
On one hand, for data-center production scenarios, using multiple GPUs for a single inference of a single model leads to a low ROI (return on investment) and can be too costly in practice.
On the other hand, this conventional approach requires extra cross-device communication, further exacerbating the efficiency problem.
% Moreover, for pipeline parallelism that slices tensors at the batch dimension for pipelining, the MatMul (matrix-multiplication) becomes even more skinny in decoder models
% and thus less efficient (we will discuss the performance problem of skinny in Sec.\ref{section: DesignOpportunity}).
% Note that the mainstream large generative models are decoder-only architecture.
GPU memory offloading and swapping is another approach to support large weights given limited GPU memory~\cite{Flex-Gen, DeepSpeed-Inference}.
However, the offloading and swapping approaches usually result in a long inference latency and thus can be impractical for online inference services.

The weight pruning methods~\cite{Model_Compression} (sparsification) have been demonstrated to be effective in reducing memory usage and computations for model inference while retaining good accuracy through removing a portion of less salient connections in neural networks.
%There are studies indicating that larger models are more robust to pruning in terms of model accuracy~\cite{Training_Big_Then_Compress}.
In practice, unstructured pruning typically retains better accuracy than more restrictive structured pruning~\cite{Wanda, Model_Compression, Deep_Compression, gomez2019learning, ullrich2017soft, SparseGPT, SparsityMethod}.
However, it is difficult to support unstructured sparsity on modern GPU architectures efficiently, especially since the unstructured sparse MatMul (SpMM) is hard to support on the high-performance but highly-structured tensor core architecture. Thus, this design direction has been largely neglected so far since higher performance is very difficult to achieve.  
For example, the state-of-the-art unstructured SpMM implementations (e.g. cuSPARSE~\cite{cuSPARSE}, Sputnik~\cite{Sputnik}) can not even outperform the dense counterpart (cuBLAS~\cite{cuBLAS}) until the model sparsity is higher than 98\% and 86\%, respectively.
Note that tensor cores on modern GPUs usually can achieve nearly an order of magnitude higher peak performance than SIMT cores.
% If we could support unstructured SpMM on Tensor Core efficiently, we could support large generative model inference with good accuracy, efficiency and cost.

To address this critical issue that bottlenecks LLM inference performance and cost, we propose \SYS{}, an efficient GPU library to support unstructured sparsity on tensor cores for large generative model inference.
With unstructured sparsity, \SYS{} addresses the memory footprint problem which leads to lower costs while retaining high model accuracy.
By being able to leverage tensor cores' high peak performance, \SYS{} achieves lower latency for unstructured SpMM compared to the existing sparse/dense MatMul solutions.
The high-level design insight of \SYS{} is this \textit{Load-as-Sparse and Compute-as-Dense} strategy.
%The key SpMMs in generative model inference are usually very \textit{skinny}.
%We make an important observation that the skinny MatMul computations are bound at global memory access (or memory bandwidth) rather than the computation capability of tensor cores.
We make an important observation that the key MatMuls in generative model inference are usually very \textit{skinny}.
Furthermore, the performance of these skinny MatMuls is bound by global memory access (or memory bandwidth) rather than the computation capability of tensor cores.
Because of this, we propose an innovative approach to support unstructured sparsity on tensor cores by leveraging sparse memory load to improve the limited memory bandwidth while effectively tolerating redundant tensor-core computation (Sec.\ref{section: DesignOpportunity}). 
%Specifically, \SYS{} reads the model weights in sparse format from off-chip memory, extracts it to on-chip memory into dense format, and feeds it to Tensor Core.
%Note that skinny MatMuls are bounded by off-chip memory access and thus the redundant computations can be tolerated (details in Sec.\ref{section: DesignOpportunity}).

Given the insight above, it is still challenging to actually design and implement this high-level \textit{Load-as-Sparse and Compute-as-Dense} approach.
First, it requires a well-designed data format for efficient sparse data storage and extraction.
The sparse data extraction is non-trivial, which requires a sophisticated design to load and extract sparse data with minimal access cost in the hierarchical GPU memory given limited on-chip memory resources.
It also introduces new challenges in designing the MatMul computation pipeline beyond conventional dense MatMul strategies.
In \SYS{}, we propose a new sparse format called (\textit{Tiled-CSL}) to support the tile-by-tile SpMM execution with tensor cores (Sec.\ref{subsubsec:Tiled-CSL}).
Based on \textit{Tiled-CSL}, we then design the \textit{sparse-to-dense transformation} approach carefully by using the distributed registers and shared memory buffers for sparse data extraction (Sec.\ref{section:DesignOverview}). 
Then, an efficient \textit{two-level overlapping strategy of memory and computation} is introduced to coordinate the \textit{sparse-to-dense transformation} on weights, the dense feature map data loading, and the tensor core operations with a full software pipeline (Sec.\ref{subsec:pipeline-design}).
% along with the \textit{minimum range of synchronization and memory barriers} to ensure the correctness while keeping the overlapping (Sec.\ref{subsubsec:sync-and-barrier}).
Finally, we propose an \textit{ahead-of-time sparse data reordering} approach to further reduce shared memory bank conflicts\footnote{
GPU shared memory is divided into multiple memory banks that can be accessed simultaneously.
Bank conflict means multiple addresses of a memory request map to the same memory bank, causing serialized accesses.}
(Sec.\ref{subsection:Ahead-of-time-Reordering}).
In summary, this paper makes the following contributions:

\begin{itemize}[noitemsep, leftmargin=*, topsep=0pt]

\item We propose \SYS{}, the first cost-effective and highly-efficient software framework for supporting large generative model inference,
%\cb{\marginpar{\cb{R3.D1}}(especially whose dense format does not fit into GPU memory)}, 
opening up the scope of enabling unstructured sparsity exploration on high-performance tensor cores. 

\item We propose a general \textit{Load-as-Sparse and Compute-as-Dense} approach to reduce memory footprint and increase the efficiency of the key skinny MatMuls by leveraging the insight of addressing the major memory bandwidth bottleneck in LLM inference while tolerating redundant tensor-core computations.

\item We propose an efficient software pipeline design to enable \SYS{} by effectively leveraging our new sparse format, the sparse-to-dense transformation, and a two-level overlapping strategy.

%leveraging on-chip resources for sparse data extraction and coordination, dense feature-map data loading, and tensor core computation effectively.
% In addition, it proposes a new sparse format \textit{Tiled-CSL} along with the \textit{ExtractRegister2Shared} and \textit{NonZero Reorder} techniques for efficient on-chip extracting.

\item \SYS{} is implemented and integrated into FasterTransformer for ease-of-use. Extensive evaluation results have shown that (1) at the kernel level, \SYS{} outperforms the state-of-the-art solutions Sputnik and SparTA by an average of $2.9 \times$ and $1.5 \times$, respectively. (2) At end-to-end framework level on OPT-30B/66B/175B models, for \textit{tokens per GPU-second}, \SYS{} achieves up to $3.8 \times$ and $3.6 \times$ improvement over DeepSpeed and FasterTransformer, with significantly lower inference cost.
% supports large generative model inference with higher inference throughput with lower cost compared to standard DeepSpeed and FasterTransformer under 80\% unstructured sparsity, 
% from $1.41\times$ to $3.8\times$ for pruned OPT-30B, OPT-66B, and OPT-175B as for \textit{tokens per GPU-second}.
% by $3.6\times$/$1.4\times$, $3.0\times$/$1.4\times$, and $2.0\times$/$1.6\times$ under 70\%, 80\%, and 90\% sparsity respectively on average.
% Besides, our sparse kernel can also outperform the state-of-the-art dense kernels cuBLAS with tensor core enabled by  $1.4\times$, $1.8\times$, and $2.2\times$.
% \item It also integrates \SYS{} kernels into FasterTransformer for end-to-end evaluation of the popular large generative models.

% The experiments on pruned OPT-30B, OPT-66B, and OPT-175B show that inference systems equipped with \SYS{} can achieve large generative model inference with higher inference throughput (from $1.41\times$ to $3.8\times$) compared to standard DeepSpeed and FasterTransformer under 80\% unstructured sparsity.
%For the OPT-30B model, our system achieves $3.4 \times$ and $3.3 \times$ higher token generation throughput than DeepSpeed (DS) and FasterTransformer(FT) with a single GPU.
%Compared to DS/FT with two GPUs used, our system achieves $1.91 \times$/$1.75 \times$, $1.87 \times$/$1.70 \times$, $1.67 \times$/$1.55 \times$, and $1.54 \times$/$1.41 \times$ higher throughput at batch sizes 8, 16, 32, and 64 respectively.
%Similar results are obtained for OPT-66B and OPT-175B models.

\end{itemize}

\section{Background}
\label{sec:background}

\begin{figure}
    \subfloat[Auto-regressive Token Generation 
    \label{fig:auto-regressive-generation}]{
        \includegraphics[width=0.48\columnwidth]{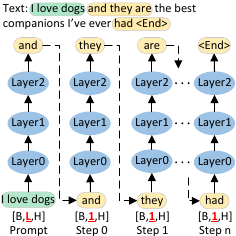}
    }
    \subfloat[KV Cache within Layer-0\label{fig:kv-cache}]{
        \includegraphics[width=0.48\columnwidth]{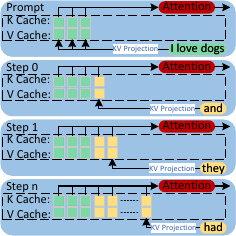}
    }
    % \vspace{-3ex}
    \caption{(a) Generative model inference; (b) KV-Cache.}
    \label{fig:opt-arch-and-kv}
\end{figure}

\subsection{Generative Model Inference}
\label{sec:GenerativeModelInference}

\textbf{\underline{Inference Procedure of Modern Generative Models.}} 
Modern generative models' inference is typically conducted in two phases: \textit{prompt processing} and \textit{token generation}. 
As illustrated in Fig.\ref{fig:auto-regressive-generation}, the generative model first performs \textit{prompt processing} to process user input sequences (\textit{'I love dogs'}) and generates the first new token (\textit{'and'}).
Then the model turns into the \textit{auto-regressive token generation} phase, where the single output token generated in step \textit{i-1} will be taken as input to generate the new token in step \textit{i} iteratively.
%This progress will terminate when meeting the \textit{'<End>'} token or reaching a predefined max output sequence length.

In the \textit{prompt processing} phase, multiple tokens within the input sequence will be processed at the same time, resulting in input tensors with shape $[B, L, H]$ (Fig.\ref{fig:auto-regressive-generation}).
$B$, $L$, and $H$ indicate the \textit{inference batch size}, \textit{prompt sequence length}, and the \textit{model hidden dimension}, respectively.
Whereas in the \textit{token generation} phase, only the single token generated in the last iteration will be taken as input thus forming the input tensors in the shape of $[B, 1, H]$.
Note that to prevent re-computations on K and V vectors
\footnote{K and V vectors of previous tokens are needed by the \textit{attention}~\cite{Attention_Is_All_You_Need} mechanism.}
of the previous tokens, a pre-allocated memory buffer (a.k.a. \textit{KV-Cache} shown in Fig.\ref{fig:kv-cache}) is usually used during token generation.
At each step, a new KV pair is generated (in yellow) and written to the \textit{KV-Cache}.

%For each layer, it also takes the intermediate activation (K and V vectors) generated in previous steps as input due to the \textit{attention}~\cite{Attention_Is_All_You_Need} mechanism. 
%To prevent re-computations on K and V vectors of the previous tokens, a pre-allocated memory buffer (a.k.a. \textit{KV-Cache}) is usually used to store \textit{K} and \textit{V} values of all the previous tokens.
%We illustrate the \textit{KV-Cache} update procedure in Fig.\ref{fig:kv-cache}.
%At each step, a new KV pair is generated (in yellow) and written to the \textit{KV-Cache}. 
%Then the \textit{Attention} computation attends to all \textit{KV} values corresponding to all the previous tokens.

\begin{figure}
    \centering
    \includegraphics[width=0.99\columnwidth]{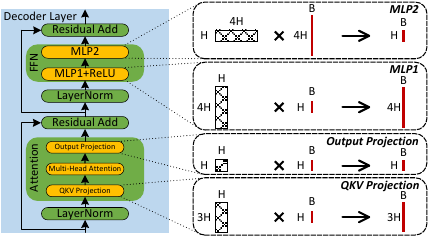}
    % \vspace{-3ex}    
    \caption{Decoder Layer Architecture. The H here means the hidden dimension aka. model dimension, which equals 12K for GPT-3. The B refers to the inference batch size which is typically small for real-time inference, e.g. 8, 16 or 32.}
    \label{fig:Decoder-Layer-Architecture}
\end{figure}

\textbf{\underline{Inference Performance Hotspot of LLMs.}} Fig.\ref{fig:Decoder-Layer-Architecture} illustrates the typical decoder architecture of a single layer in modern attention-based generative models.
% Among the operators in the decoder computation graph, matrix multiplication (MatMul) takes a majority part of end-to-end latency.
There are in total four major matrix multiplications (or \textit{MatMuls}) in the decoder layer: \textit{QKV Projection}, \textit{Output Projection}, \textit{MLP1}, and \textit{MLP2}.
Unlike previous encoder-centric non-generative language models (e.g., BERT~\cite{BERT}) of which the performance bottleneck is mainly at the multi-head attention computation, 
generative model inference's performance is heavily bounded by these four MatMuls. 
According to our experiments on OPT-66B~\cite{OPT-Models} model inference, these four MatMuls are the top contributors to the overall latency which accounts for $76.8\%$ of the end-to-end execution time and also the top contributor of the memory consumption which accounts for $83.8\%$ of the overall memory usage.

\subsection{ Matrix Multiply in LLM Inference}
\label{sec:MatrixMultiply}

\textbf{\underline{Skinny Matrix Multiply.}}
% \cb{
% \marginpar{\cb{R1.D4}}
The Matrix Multiply (MatMuls) in Fig.\ref{fig:Decoder-Layer-Architecture} can be formalized as $C = A \times B$, where $A$ is the weight matrix of shape $[M, K]$ and $B$ is the feature map matrix of shape $[K, N]$.
In this paper, we call these MatMuls \textit{"Skinny MatMuls"}, as their $N$ dimensions are much smaller than the $M$ and $K$ dimensions.
\footnote{For these MatMuls, $M$ and $K$ are integer multiples of hidden size while $N$ equals inference batch size (typically orders of magnitude smaller than hidden size).}
% }

\textbf{\underline{Differences between Tensor/SIMT cores.}}
% \cb{
% \marginpar{\cb{R1.D2}}
SIMT cores (aka. CUDA cores) are general-purpose execution units that handle a wide range of instructions for parallel execution, while tensor cores~\cite{Ampere_WhitePaper, Hopper_WhitePaper} are specialized units designed specifically to accelerate dense MatMul computations.
Tensor cores provide significant acceleration for dense MatMuls, e.g. $16\times$ higher throughput than SIMT cores in A100 GPUs with FP32 accumulation.
% }

% \cb{
% \marginpar{\cb{R1.D2}}
Conventional techniques leveraging SIMT cores for sparse MatMuls can not be directly applied to tensor cores as SIMT and tensor cores work at very different granularity.
SIMT cores work on the granularity of scalar values, e.g. the \textit{FMA} instruction on scalar value.
The per-element granularity makes it easy to do computation skipping at the element level for SpMM.
% If the Operand1 is zero (the corresponding scalar element within the weight matrix is pruned), SIMT-based Sparse MatMul kernels can simply skip executing this FMA instruction.
However, tensor cores work at a much more coarse-grained granularity than SIMT cores, e.g., perform a matrix multiply between two matrices with the shapes of $16\times16$ and $16\times8$ in each instruction.
Thus, tensor cores do not allow skipping arbitrary element-level computations.
% , as they can only take small but dense matrices as inputs and output a dense matrix.
% }

\section{Opportunities and Insights}
\label{sec:motivation}

\subsection{Unstructured Sparsity on Tensor Cores}
\label{section:Demand-of-Unstructured-Sparsity}

% \cb{
% \marginpar{\cb{R1.D1}}
There are two typical types of pruning principals.
The most flexible pruning strategy (\textit{unstructured sparsity}) is to remove less salient elements without considering the distribution of the pruned elements in the weight matrix.
Taking \textit{magnitude pruning} for example, we can rank all the elements in the matrix based on their absolute values and then remove the weights with the smallest magnitude.
Another strategy (\textit{structured sparsity}) is to prune the less salient weights, but at the same time, some kind of structural criteria must be enforced.
For example, the weight matrices can be split into non-overlapping $8\times1$ vectors \cite{TC-StructuredSparsity, TC-Quantized} or $32\times32$ blocks \cite{Blocked-ELL}, and then each vector/block is either kept or removed during pruning.
% }

% \cb{
% \marginpar{\cb{R1.D1}}
In short, the major difference between structured and unstructured pruning is that extra constraints must be satisfied for structured pruning.
Even though structured sparsity is friendly for hardware acceleration, it suffers from more severe model accuracy degradation \cite{Wanda, Model_Compression, Deep_Compression, gomez2019learning, ullrich2017soft, SparseGPT} as it limits the freedom of deciding which element to prune.
As shown in~\cite{SparsityMethod}, compared to structured sparsity which has 5\% accuracy drop, unstructured sparsity only results in 1\% accuracy drop.
In our experiments, the OPT-like models could greatly preserve accuracy through retraining-based unstructured pruning~\cite{liu2018rethinking, han2015learning} at 80\% sparsity (e.g., the accuracy only decreases from 85.55\% to 84.11\% for OPT-30B).
% }

\begin{figure}
    \centering
    \includegraphics[width=\columnwidth]{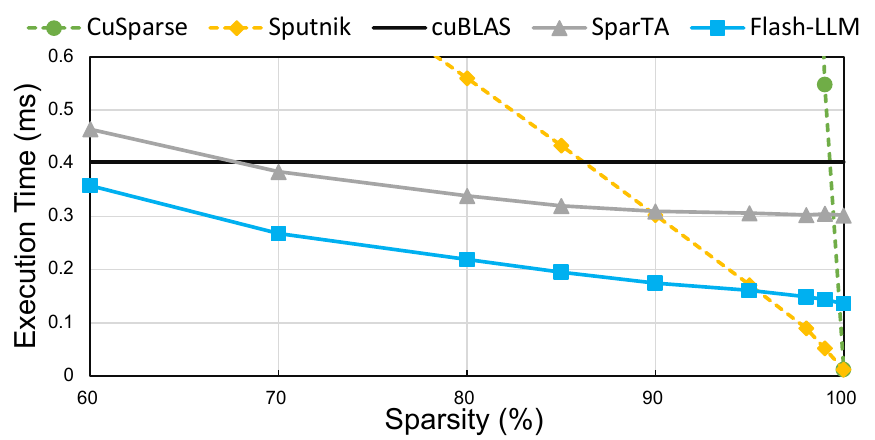}
    % \vspace{-6ex}  
    \caption{Performance of an unstructured SpMM (M/K/N =hidden\_size*4/hidden\_size/batch\_size=36K/9K/8) under different designs on GPU.
    SIMT core centric designs are indicated with dash lines while tensor core centric designs are indicated with solid lines (including our solution Flash-LLM).
    }
    \label{fig:ExistingSparseGPUKernels}
\end{figure}

However, the conventional techniques for supporting random unstructured sparsity in SpMM execution are not effective since they focus on leveraging SIMT cores without a sophisticated way of utilizing high-performance tensor cores. 
Fig.\ref{fig:ExistingSparseGPUKernels} shows the performance comparison of different techniques for SpMM on an OPT-66B inference task with batch size 8. 
Note that the standard pruning for LLM inference typically requires a moderate level of sparsity (e.g., 80\%) to preserve model quality while reducing memory footprint.
CuSparse~\cite{cuSPARSE}, the NVIDIA SpMM library, shows poor performance as it is mainly designed for scientific applications where matrices are extremely (99\%+) sparse.
Sputnik~\cite{Sputnik}, the state-of-the-art SIMT-core-centric optimization for unstructured SpMM on DL tasks still cannot outperform cuBLAS(dense) until a high sparsity is reached.

% \cb{
% \marginpar{\cb{R3.D3}}
We observe that sparse MatMul kernels in existing sparse libraries are usually slower than their dense counterpart (cuBLAS\cite{cuBLAS}).
The reason is that cuBLAS have leveraged the tensor cores, while sparse MatMul kernels are leveraging SIMT cores in state-of-the-art solutions.
Note that A100 GPU \cite{Ampere_WhitePaper} can provide $16\times$ higher computational throughput using tensor cores than using SIMT cores for dense MatMuls.
Although Sputnik can effectively leverage SIMT cores for unstructured sparsity processing, its performance is still limited by the peak performance of SIMT cores.
% }

% \cb{
% \marginpar{\cb{R1.D1}}
Due to the clear peak performance discrepancy between SIMT and tensor cores, there is a strong demand for high-performance unstructured SpMM support for LLM inference. 
However, \textit{it is non-trivial to enable high-performance unstructured SpMM onto the highly restrictive tensor cores as tensor cores do not allow skipping arbitrary scalar-level computations (described in section \ref{sec:MatrixMultiply})}.
% }
Previous SpMM works are based either on highly structured sparse matrices~\cite{TC-StructuredSparsity, Shfl-BW, Blocked-ELL} (not random unstructured sparsity), or for extremely high sparsity ratio~\cite{TC-GNN} (i.e., >95\%, inappropriate for LLMs' inference accuracy), rather than random unstructured sparsity at a moderate sparsity ratio range for high accuracy.
SparTA\cite{SparTA} leverages sparse tensor cores~\cite{SparseTensorCore-NVIDIA} for major computations.
However, it cannot effectively exploit high sparsity as sparse tensor cores only support 50\% sparsity (i.e., 2:4 sparsity).
As shown in Fig.\ref{fig:ExistingSparseGPUKernels}, the performance of SparTA is lower than \SYS{} especially as the sparsity increases.
%However, it cannot effectively exploit a large sparsity range available as sparse tensor cores only support 50\% sparsity (i.e., 2:4 sparsity).
%If the sparsity available is higher than 50\%, SparTA has to pad zeros to the sparse matrix which may result in non-negligible performance losses. Furthermore, when the non-zeros still cannot satisfy the 2:4 requirement, another SIMT kernel needs to be launched for the corresponding computations (further computation splitting), resulting in additional overhead.

\subsection{Design Opportunities}
\label{section: DesignOpportunity}

Given the natural mismatch between the unstructured SpMM computation and the highly structured tensor core architecture, it is essential to have a highly efficient SpMM solution on tensor cores according to the workload characteristics of modern LLMs.
As discussed in Sec.\ref{sec:MatrixMultiply}, MatMuls in modern LLMs inference are skinny.
As a result, the bottlenecks of the skinny MatMul computations are the off-chip memory access and bandwidth limitations, rather than the arithmetic processing on tensor cores.
Based on this observation, the basic idea to address this problem is through a \textit{Load-as-Sparse and Compute-as-Dense} approach.
Specifically, the GPU kernel loads the weight matrices from global memory in sparse format with reduced size and reconstructs the corresponding dense format in high-speed on-chip buffers for tensor core computation.
% Then, the reconstructed dense matrices will be fed to Tensor Cores for computation.
The key insight is that the bottleneck for LLM inference is not at the computation side thus we can tolerate the redundant computations with tensor cores.
We systematically describe how the off-chip memory transactions become the performance bottleneck in Sec.\ref{subsubsec:skinny-matmul-bottleneck}, and why it is possible to tolerate such redundant computation for skinny SpMMs in LLMs' inference in Sec.\ref{sec:sparse-load-dense-compute}.

\subsubsection{Performance Bottleneck of Skinny MatMuls in LLM Inference.}
\label{subsubsec:skinny-matmul-bottleneck}

\begin{figure}
    \centering
    \includegraphics[width=\columnwidth]{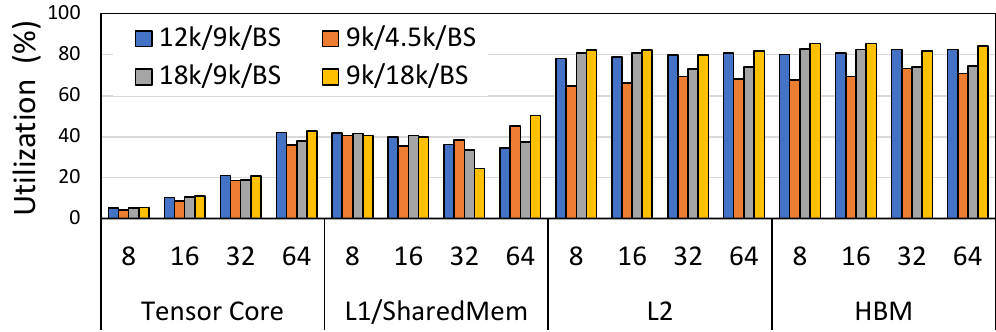}
    % \vspace{-5ex}
    \caption{GPU utilization Breakdown. The MatMuls profiled in this figure are the most time-consuming parts during OPT-66B inference (with 2 GPUs) at batch sizes 16, 32, 64, and 128.}
    \label{fig:GPU_UtilizationBreakdown}
\end{figure}

We analyze the performance bottleneck of skinny MatMuls execution starting from dense MatMul workloads in LLMs.
According to our profiling results for OPT-66B~\cite{OPT-Models}, the average utilization of tensor cores is around 5.0\%, 10.1\%, 19.9\%, and 39.7\% under typical batch sizes of 8,16,32 and 64 as shown in Fig.\ref{fig:GPU_UtilizationBreakdown}, while the bandwidth of global memory is already fully saturated.
The underlining cause for this is that the compute intensity (i.e., FLOP/Byte) of skinny MatMul is very low.
For a MatMul described in sec. \ref{sec:MatrixMultiply},
the total operations conducted are $2MNK$ floating-point operations (FLOP), and the corresponding data read is $2(MK + KN)$ bytes with FP16 data type.
Thus, the computational intensity ($CI$) is:
\begin{equation}
% reducing the space around equations
% \setlength\abovedisplayskip{0pt}
% \setlength\belowdisplayskip{0pt}
\label{equation:CI}
% \begin{split}
    CI = \frac{M \times N}{M + N}
% \end{split}
\end{equation}

According to Equation.\ref{equation:CI}, it is easy to demonstrate that 
the overall $CI$ of a MatMul can be easily restricted by \textbf{either} a small M or N dimension.
%For example, when $M$ is 1024 and $N$ is 768, $CI$ will be 439. 
For instance, for a skinny MatMul where $N$ is 16, $CI$ will have a strict upper bound of 16 no matter how big the $M$ dimension is.
Note that in generative LLM models, the $N$ dimension equals the inference batch size which is usually very small in production environments.
Thus, the $CI$ is strongly bounded by the $N$ dimension in real-time LLM inference.
According to the roofline model~\cite{Roofline_Model}, the performance of a kernel with low computational intensity will be easily bounded by memory bandwidth.
%Additionally, the skinny MatMul leads to low data reuse on the L2 cache.
%The $N$ dimension (batch size) of the MatMul in Fig.\ref{fig:Decoder-Layer-Architecture} is quite small for LLM inference, there will be only one thread-block tile at the N dimension in the tiling-based MatMul implementation.
%This means the data of the left-hand-side matrix will not be reused between different thread-block tiles, leading to no L2 cache locality.
%As a result, the skinny MatMuls rely on the low bandwidth of global memory for data fetching to the on-chip memory under the low compute intensity. 
%Additionally, the negative impact of the low compute intensity on performance can further exacerbate as the gap between GPU peak performance and GPU global memory bandwidth is growing bigger due to manufacturing challenges.
%For example, NVIDIA's H100 GPUs~\cite{Hopper_WhitePaper} achieves $3.2\times$ higher peak half-precision performance than A100~\cite{Ampere_WhitePaper} while H100's global memory bandwidth is only improved by $1.6\times$.

\subsubsection{Load as Sparse, Compute as Dense}
\label{sec:sparse-load-dense-compute}

%Given Equation.\ref{equation:CI}, sparsity on weight matrix will reduce both the numerator (FLOP factor) and denominator (memory load factor), still resulting in low $CI$.
%In this way, the MatMul execution is still tend to be bounded by memory load, and leaving the computation units under utilized.
%Thus the possible speedup of sparsity mainly comes from reduced memory transactions.
% Given that there are limited opportunities for skipping computations in generative models with moderate-level sparsity, s without skipping any computations can easily outperform SIMT cores with computation-skipping.
Given that the bottleneck of skinny MatMul comes from memory access/memory bandwidth rather than arithmetic computation, we propose the basic idea of \textit{Load-as-Sparse and Compute-as-Dense} here which leverages the performance boost from reduced memory access while enabling the efficient use of tensor cores for unstructured sparsity (refer to Sec.\ref{sec:DesignMethodology} for details).
Under this basic idea, given the sparsity ratio $\beta$ (the weight matrix $A$ is sparse while the feature map matrix $B$ is dense), the computational intensity can be improved to:
\begin{equation}
\label{equation:CI_Sparse}
% \begin{split}
    CI_{SparseLoad}\footnote{
    It's worth noting that we do not take sparse index overhead into the theoretical consideration here. In practice, the real CI would be a bit lower than this equation.} 
    = \frac{M \times N}{M \times (1 - \beta) + N}
% \end{split}
\end{equation}

\begin{figure}
    \centering
    \includegraphics[width=0.7\columnwidth]{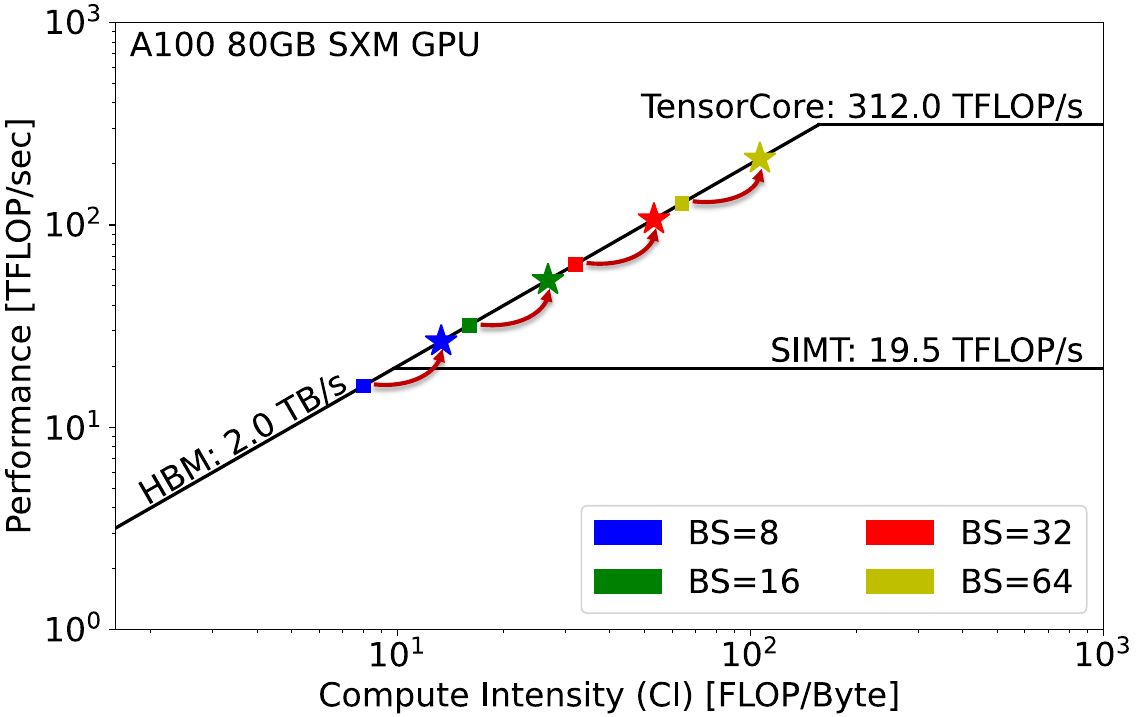}
    % \vspace{-2ex}
    \caption{Roofline model for skinny MatMuls. 
    The solid Squares refer to the CI and the  performance upper bound 
    for dense solutions (e.g. cuBLAS), while the solid Stars represent the improved CI and the new performance bound with our \textit{Load-as-Sparse Compute-as-Dense.}
    Note that the vertical axis is displayed on a logarithmic scale.
    } 
    \label{fig:RooflineModel}
\end{figure}

Fig.\ref{fig:RooflineModel} shows the CIs and their corresponding achievable tensor core performance of a typical MatMul ($M$: 48k, $N$: BS, $K$: 12k) in OPT-175B model inference with different batch sizes.
According to the figure, MatMuls in generative model inference with different batch sizes all face memory wall issues.
As a result, the dense MatMuls kernels can only achieve 5.1\%, 10.3\%, 20.5\%, and 40.1\% peak performance of tensor cores bounded by insufficient global memory bandwidth.
These theoretical values are consistent with our actual measurements in Fig.\ref{fig:GPU_UtilizationBreakdown}.
In theory, with \textit{Load-as-Sparse and Compute-as-Dense} approach under 40\% sparsity, the tensor cores utilization can be improved to 8.5\%, 17.1\%, 34.2\%, and 68.2\%.

\section{Design Methodology}
\label{sec:DesignMethodology}

% Given the clear insight in Sec.\ref{sec:motivation}, it is still challenging to realize the \textit{Load-as-Sparse and Compute-as-Dense} approach.
% It first requires a well-designed data format for efficient sparse data storing and extraction.
% The sparse data extraction (i.e., reconstructing the dense data on-chip from sparse data on global memory) is non-trivial, which brings extra data movements and computations.
% With limited on-chip memory resources, it requires a sophisticated design to load and extract sparse data with minimal access cost in the hierarchical GPU memory.
% It also introduces new challenges of designing the MatMul computation pipeline beyond conventional dense MatMul.

\SYS{} leverages both SIMT cores and tensor cores effectively for efficient unstructured SpMM computation.
The flexible SIMT cores are exploited for \textit{Sparse-to-Dense Transformation} (i.e., \textit{Load-as-Sparse}) while tensor cores are used for compute-intensive tensor computations (i.e., \textit{Compute-as-Dense}).
We will give an overview of the high-level optimizations of \SYS{} in Sec.\ref{section:DesignOverview}.
Then we will describe the design of \SYS{}'s computation pipeline in Sec.\ref{subsec:pipeline-design}.
We illustrate the novel sparse format (\textit{tiled-CSL} format) and the efficient memory access techniques in Sec.\ref{subsec:sparse-encoding-and-runtime-parsing}.
Finally, we described the end-to-end inference system enabled by our \SYS{} in Sec.\ref{section:System-Integration}.
% Note that for the SpMM computation $C = A \times B$, we regard $A$ as the sparse weight tensor and $B$ as the dense feature map tensor in our discussion.
% Our SpMM kernel \SYS{} implements the computation $AB \rightarrow C$, where A is the sparse matrix stored in Tiled-CSL format.
% In the following sections, we refer to matrices A, B, and C as the sparse matrix, dense matrix, and output matrix respectively.
% We design the \textit{Tiled-CSL} format for $A$ to support the efficient data loading and computation (details in Sec.\ref{section:Tiled-CSL}).
% \SYS{} supports matrix A is stored with our customized sparse format Tiled-CSL described in Section.\ref{section:Tiled-CSL}, while B and C are stored in column-major dense format.

\subsection{Design Overview}
\label{section:DesignOverview}

% As discussed in section \ref{section: DesignOpportunity}, global memory transaction is the main bottleneck for GPU kernel performance.
% When mapping the \textit{Sparse-to-Dense Transformation} to the existing GPU memory hierarchy, our first priority is to minimize its bandwidth overhead toward global memory.
% The straightforward solution is to load the weight matrices in sparse encoding from global memory, re-construct the dense matrices and write them back to global memory.
% The dense matrices are then loaded from global memory for normal MatMul computations.
% However, the total global memory data movements will be $2\times$ more compared to the dense kernel.
% We aim to access the weight matrix in global memory only once in our kernel design.
% The sparse encoding of the sparse matrix will be loaded from global memory to reconstruct the dense format on GPU's on-chip buffers, without writing the corresponding dense matrices back to global memory.
% However, it is clear that GPU's on-chip buffer can not accommodate the whole sparse matrix.

\begin{figure*}
    \centering
    \subfloat[Load-as-Sparse, Compute-as-Dense. 
    \label{fig:LSCD}]{
        \includegraphics[width=0.3\linewidth]{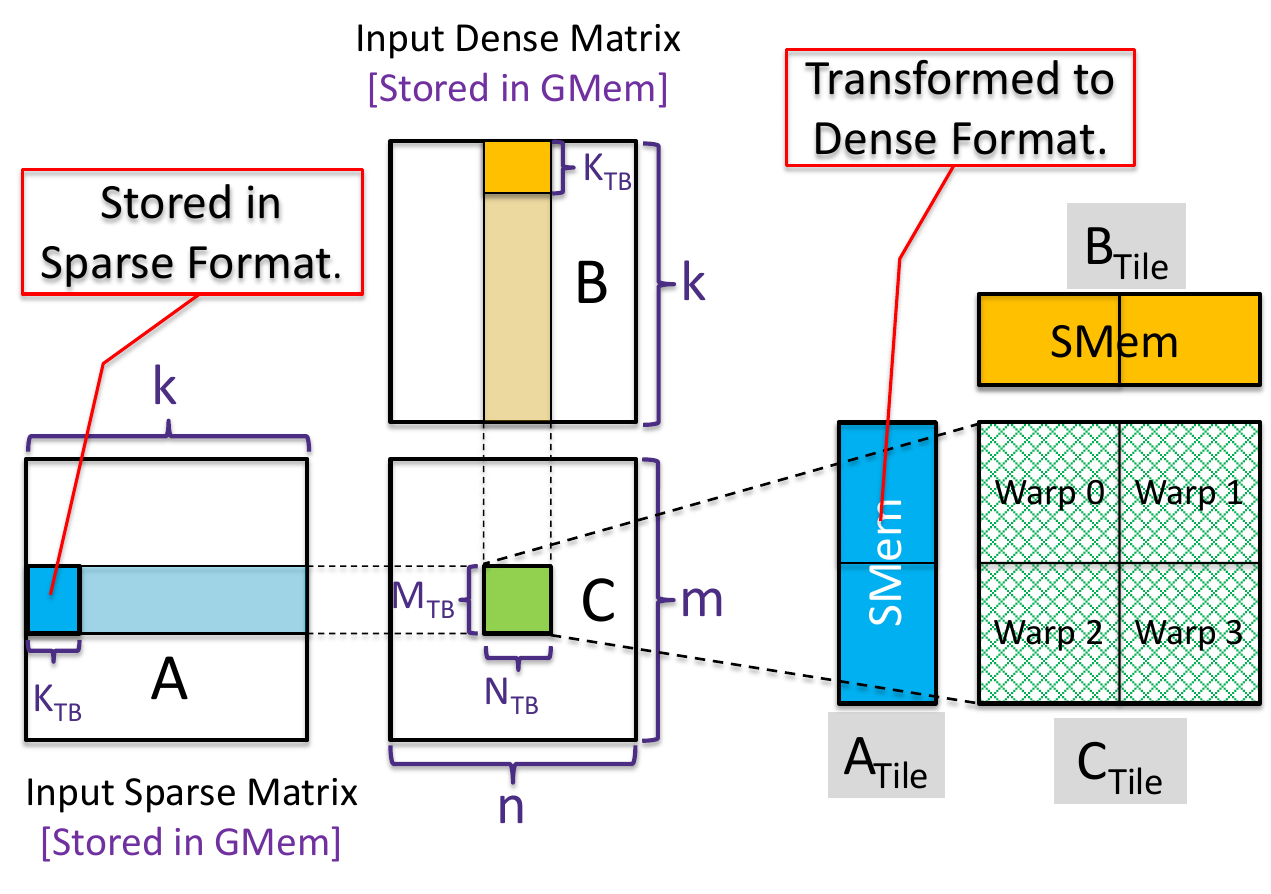}
    }
    \subfloat[Sparse-to-Dense Transformation. \label{fig:Sparse-to-Dense-Transformation}]{
        \includegraphics[width=0.27\linewidth]{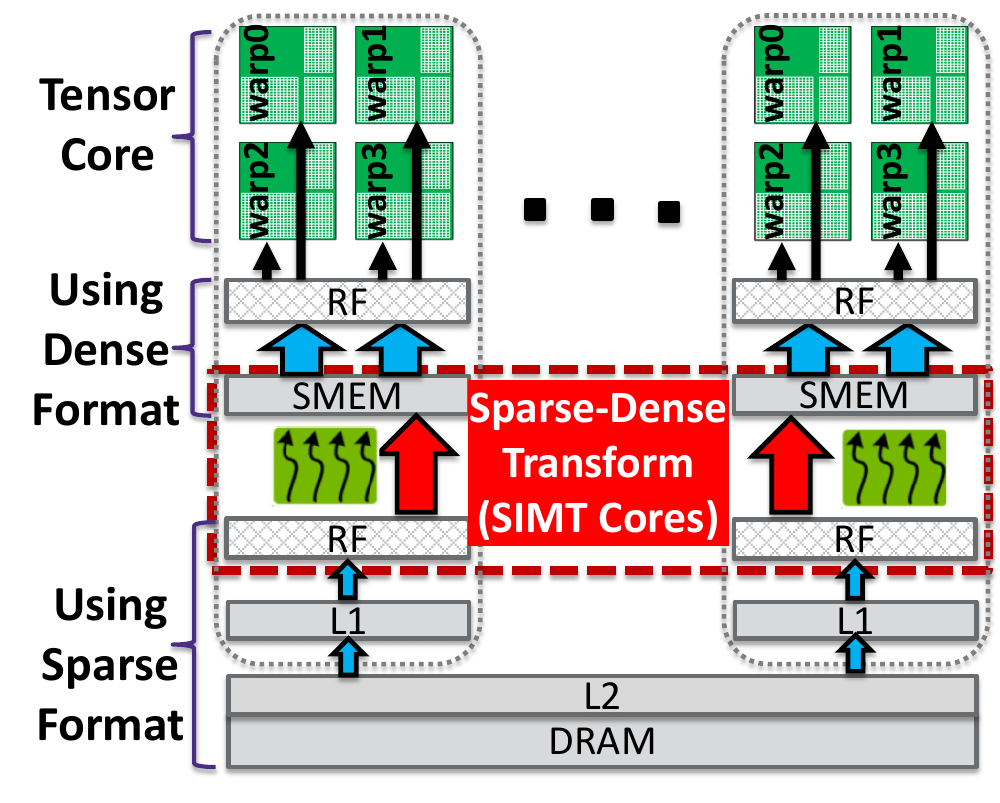}
    }
    \subfloat[Pipelined memory and tensor core operations. \label{fig:DataTransfer_Computation_Pipeline}]{
        \includegraphics[width=0.37\linewidth]{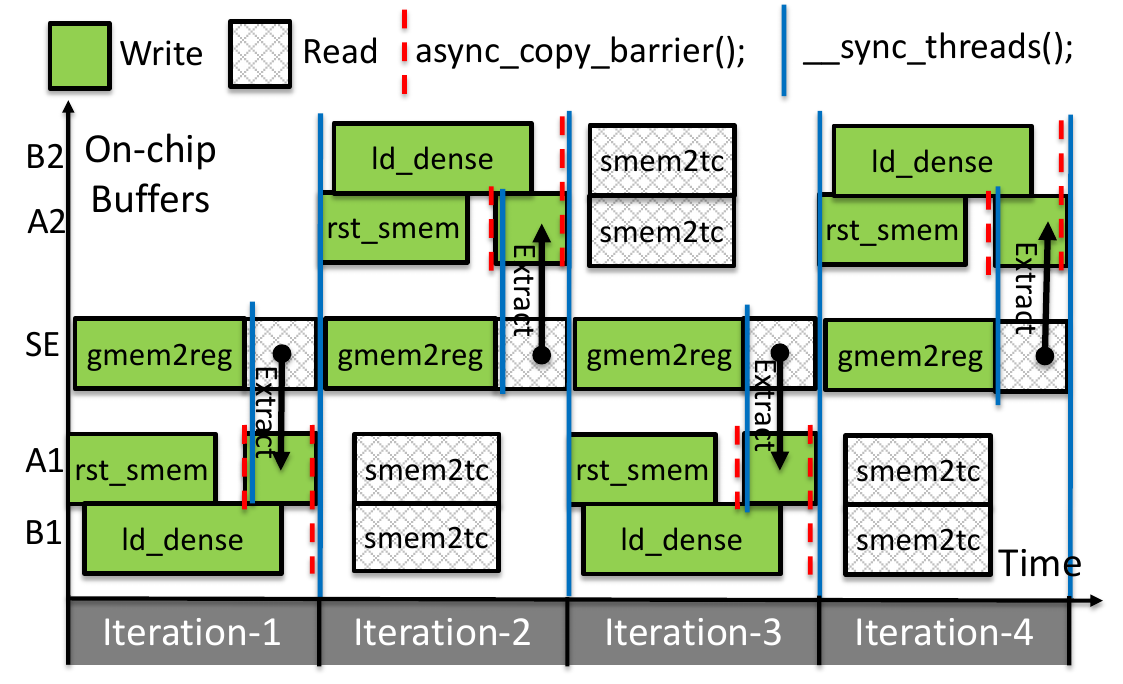}
    }
    \vspace{-2ex}
    \caption{Design Overview.}
    \label{fig:Design-Overview}
\end{figure*}

We use the tiling-based approach for the SpMM computations in \SYS{}.
% As shown in Fig.\ref{fig:LSCD}, each thread-block processes a tile of $C$ matrix.
% The $A$ and $B$ matrices are loaded and computed tile-by-tile at $K$ dimension.
% Specifically, we use the \textit{Tile-By-Tile Transformation} for the sparse $A$ matrix, where one tile of $A$ will be transformed from sparse encoding to dense format each time.
% It means that we only transform the sparse tile into a dense format when it is about to be consumed by the Tensor Core operations.
% We adopted the tiling strategy of dense MatMul to divide our weight matrix into smaller tiles and conduct the tile transformation on demand.
% It means that we only transform the sparse tile into a dense format when it is about to be consumed by the Tensor Core operations.
Fig.\ref{fig:LSCD} shows the tiling method of \SYS{}, where each thread block (TB) is in charge of calculating a tile (e.g., the green tile in the shape of $M_{TB}*N_{TB}$) in output matrix $C$.
For each iteration, each thread block loads $A_{Tile}$ (shape $[M_{TB}, K_{TB}]$) in sparse and $B_{Tile}$ (shape $[K_{TB}, N_{TB}]$) in dense from global memory.
$A_{Tile}$ is then transformed to dense format with \textit{Sparse-to-Dense Transformation} shown in Fig.\ref{fig:Sparse-to-Dense-Transformation} and stored in shared memory while $B_{Tile}$ is directly stored in shared memory.
Finally, each thread block consumes the dense data in shared memory and generates the output tile through tensor core computations.

Fig.\ref{fig:Sparse-to-Dense-Transformation} shows the overall kernel behavior of \SYS{} from the microarchitecture aspect.
Shared memory is used as the workspace to store the extracted data from sparse to dense format, where all threads within the thread block work together collaboratively to load sparse encoding (SE) \footnote{
We refer to the data of $A$ in the sparse format as \textit{sparse encoding} in this paper.}
of $A_{Tile}$ from global memory and store them in shared memory with the dense format.
Note that each tile on global memory is encoded into sparse encoding with fewer bytes than the dense format, 
for which the corresponding global memory data movements can be reduced compared to dense MatMul.
Specifically, the basic idea of \textit{Sparse-to-Dense Transformation} is extracting non-zero elements from the sparse encoding on global memory to their corresponding locations in the dense format on shared memory while other locations are filled with zeros.
We use the distributed registers as the intermediate buffer to store the non-zero elements before extracting them to shared memory.
% The sparse encoding on the distributed registers will be stored to shared memory collaboratively by the threads within the thread block.
We do not use shared memory as this intermediate buffer to avoid the turn-around shared memory access of the sparse encoding. 
% The Sparse Encoding is staged directly using distributed registers instead of using shared memory to eliminate the turn-around shared memory access of sparse encoding.
% Otherwise, the sparse encoding will first be stored in shared memory, and then be loaded by SIMD cores for the non-zero extraction.
%Note that there are enough registers not used in the MatMul implementation, for which there are enough registers as intermediate high-speed buffers for sparse encoding.
% Although each thread can have no more than 256 registers in state-of-the-art A100 GPUs, the aggregated capacity of registers (256KB) is usually bigger than that of shared memory (at most 164KB).
% What's more, we found that there are enough registers not used in existing dense kernels for LLM inference, which can be used as temporary high-speed buffers.

%Registers can not be used as workplace for this transformation due to their limited visibility as described in section \ref{Section:}. \cb{[Should be analyzed in detail somewhere]}

\subsection{Computation Pipeline Design of \SYS{}}
\label{subsec:pipeline-design}

Given that each thread/thread-block consumes a large fraction of the overall registers/shared-memory as buffers for tiling, the GPU thread-level parallelism (TLP) is inherently low. 
Thus, it is important to optimize the instruction-level parallelism.
We describe the software pipeline of \SYS{} in this section where the off-chip memory loading (\textit{Sparse-to-Dense Transformation}) and tensor core computations are processed in a pipeline manner efficiently.

\subsubsection{Two-level Overlapping of Memory and Computation.}
\label{subsubsec:overlapped-execution}

%\begin{figure}
%    \centering
%    \includegraphics[scale=0.45]{./Figures/DataTransfer_Computation_Pipeline.pdf}
%    \caption{Pipelined memory and Tensor Core operations.}
%    \label{fig:DataTransfer_Computation_Pipeline}
%\end{figure}

As described in Sec.\ref{section:DesignOverview}, it requires several stages to load the sparse encoding from the global memory to shared memory in dense format for each $A_{Tile}$.
Specifically, it requires loading sparse encoding from global memory to the distributed registers (\textit{gmem2reg} stage), resetting the target shared memory buffer with zero (\textit{rst\_smem} stage), and writing the sparse encoding from registers to the corresponding locations on shared memory buffer (\textit{extract} stage).
As for $B_{Tile}$, which is already in dense format, it only requires loading directly from global memory to the target shared memory buffer (\textit{ld\_dense} stage).
Finally, the \textit{smem2tc} stage loads the shared memory data of $A_{Tile}$ and $B_{Tile}$ and executes tensor core computations.

As shown in Fig.\ref{fig:DataTransfer_Computation_Pipeline}, \SYS{} exploits a two-level overlapping of the above memory and computation stages for efficient execution.
On one hand, it leverages the software pipeline through double buffering to overlap off-chip memory loads and \textit{Sparse-to-Dense transformation} with tensor core computation, called \textit{inter-iteration overlapping}.
On the other hand, it overlaps the stages of off-chip memory load within \textit{Sparse-to-Dense transformation} for more efficient memory activities, called \textit{intra-iteration overlapping}.
% Fig.\ref{fig:DataTransfer_Computation_Pipeline} shows how the above stages are organized in a two-level overlapping manner.
The horizontal axis of Fig.\ref{fig:DataTransfer_Computation_Pipeline} represents the execution time while the vertical axis represents the activities with the double buffer.
It uses two shared memory buffers for $A_{Tile}$ (corresponds to A1 and A2) and $B_{Tile}$ (corresponds to B1 and B2), and one register buffer reused in different iterations (corresponds to SE).
Specifically, SE in Iteration-1/3 (Iteration-2/4) and A1 (A2) correspond to the \textit{Sparse-to-Dense transformation} process of $A_{Tile}$ on the first (second) set of buffer,
and B1 (B2) corresponds to the data movement of $B_{Tile}$ on the first (second) set of buffer.
As for \textit{inter-iteration overlapping}, as shown in Iteration-2 in Fig.\ref{fig:DataTransfer_Computation_Pipeline}, 
while loading data from shared memory and executing tensor core computations for data on the first set of buffers, \SYS{} loads and extracts data from global memory to shared memory for the second set of buffers.
As for \textit{intra-iteration overlapping}, the activities of A1 and B1 are processed in parallel, and the \textit{gmem2reg} and \textit{rst\_smem} stages on $A_{Tile}$ are also processed in parallel.
In this way, the sparse data loading, dense data loading, and tensor core computations can be overlapped efficiently.

A critical design for \textit{Sparse-to-Dense transformation} is explicitly using registers as data buffers between global memory and shared memory.
In \SYS{}, the sparse encoding movement from global memory to shared memory is explicitly split into two stages, i.e. LDG (loading data from global memory to registers) instructions during \textit{gmem2reg} and STS (storing data to shared memory from registers) instructions during \textit{extract} as shown in Fig.\ref{fig:DataTransfer_Computation_Pipeline}.
On one hand, the split two-stage design helps to increase instruction-level-parallelism (ILP) to hide high global memory access latency.
Note that each pair of LDG and STS instructions have load-use dependency.
If we do not split the \textit{gmem2reg} and \textit{extract} into two stages but launch each pair of LDG and STS instructions in the adjacent cycles 
(e.g. directly storing to shared memory after loading from global memory)
,
each GPU thread will execute instructions in the order of $LDG_0, STS_0, LDG_1, STS_1, ...$ without ILP of global memory load (i.e., LDGs).
By splitting the two instructions into two stages as in Fig.\ref{fig:DataTransfer_Computation_Pipeline}, the execution order will be $LDG_0, LDG_1, ..., STS_0, STS_1, ...$ and results in a high ILP of global memory load.
On the other hand, splitting the data movement into \textit{gmem2reg} and \textit{extract} enables the overlapping opportunity between \textit{gmem2reg} and \textit{rst\_smem}.
Note that STS instructions within \textit{extract} should not be launched before the completion of \textit{rst\_smem} stage, otherwise, the data written by STS instructions might be overwritten by \textit{rst\_smem} incorrectly.
%The execution of \textit{gmem2reg} and \textit{rst\_smem} can be overlapped once the LDG and STS instructions are split into two separate stages, which further increases ILP.
The execution of \textit{gmem2reg} and \textit{rst\_smem} can be overlapped once the \textit{gmem2reg} contains no shared memory write (all STS instructions are assigned to the \textit{extract} stage), which further increases ILP.

\subsubsection{Minimum Range of Synchronizations and Memory Barriers}
\label{subsubsec:sync-and-barrier}

Given the complex pipeline in Fig.\ref{fig:DataTransfer_Computation_Pipeline}, it requires a set of thread synchronizations and memory barriers to ensure correctness.
\SYS{} inserts the minimum range of synchronizations and memory barriers to ensure correctness while keeping the overlapping.

To avoid the data written by \textit{extract} being overwritten by \textit{rst\_smem} incorrectly, \SYS{} inserts the explicit thread-block level synchronization between the two stages to ensure that all the threads have finished their work resetting the A1/A2 buffer in shared memory, shown as the first yellow lines in each iteration in Fig.\ref{fig:DataTransfer_Computation_Pipeline}.
Meanwhile, it also requires another synchronization to ensure that all data movements and tensor core operations of the current iteration are completed before starting the next iteration, 
shown as the second yellow lines in each iteration in Fig.\ref{fig:DataTransfer_Computation_Pipeline}.
Take Iteration-1 as an example, we should make sure that all the threads have finished writing the A1 and B1 shared memory buffers before we start Iteration-2, as the data on the shared memory buffers will be loaded to tensor cores as inputs in Iteration-2.
Besides, we have to make sure that all the threads have finished reading the data from the register buffer for \textit{extract} in Iteration-1 before letting them be overwritten by the \textit{rst\_smem} in Iteration-2.

Beside the thread-block synchronizations, it also requires memory barriers after asynchronous copy activities of global-to-shared data movement.
\SYS{} makes use of the asynchronous copy primitives on GPU for the overlapping of data movement and other activities.
Note that the asynchronous copy primitive \textit{cp.async}, starting from NVIDIA Ampere GPU~\cite{Ampere_WhitePaper}, allows moving data from global memory to shared memory in the background asynchronously while executing other computations in the foreground.
Specifically, both the \textit{rst\_smem} and \textit{ld\_dense} stages use the \textit{cp.async} primitives to enable the overlapped execution on the double buffer.

To enable a fine-grained pipeline execution, \SYS{} uses different async-copy barriers for \textit{rst\_smem} and \textit{ld\_dense} stages.
As shown in Fig.\ref{fig:DataTransfer_Computation_Pipeline},
the \textit{extract} stage waits for the completion of only \textit{rst\_smem}, while the final thread-block barrier of each iteration waits for the completion of all previous cp.async operations.
In this way, the \textit{extract} stage could be overlapped with the \textit{ld\_dense} stages.

\begin{algorithm}\footnotesize
\caption{\SYS{} SpMM kernel pseudo code.} 
\label{alg:Kernel_Overview}
\begin{algorithmic}[1]
\State \textbf{Inputs:} \text{\textit{SparseMatrix} $A$, \textit{Matrix} $B$}
\State \textbf{Output:} \text{\textit{Matrix} $C$}

\State $Initialize\_Pipeline();$
\State $offset = subArray(A.offset);$

\State \textcolor{blue}{int} $start_{prefetch} = offset[1];$
\State \textcolor{blue}{int} $nnz_{prefetch} = offset[2] - offset[1];$ 

\For {\textcolor{blue}{int} $id = 0;$ $id < K_{Global}/K;$ $id++$}
    \State {\textcolor{brown}{//Prefetch startIdx and nnz.}}
    \State \textcolor{blue}{int} $start = start_{prefetch};$
    \State \textcolor{blue}{int} $nnz = nnz_{prefetch};$
    \State $start_{prefetch} = offset[id+2];$
    \State $nnz_{prefetch} = offset[id+3] - offset[id+2];$ 
    \State \textcolor{brown}{//Set pointers for double-buffer.}
    \State {\textcolor{blue}{half*} $smem\_w = smem + ((id+1)\%2)*OFFSET;$}
    \State {\textcolor{blue}{half*} $smem\_r = smem + (id\%2)*OFFSET;$}
    \State \textcolor{brown}{//Launch Asynchronous Memory Operations.}
    \State $InitSharedMem(smem\_w);$ \Comment{\textcolor{teal}{rst\_smem}}
    \State $cp\_async\_commit();$
    \State $CopyGlobal2Reg(A.nz+start, nnz)$ \Comment{\textcolor{teal}{gmem2reg}}
    \State $CopyGlobal2Shared(smem\_w, B.data)$ \Comment{ld\_dense}
    \State $cp\_async\_commit();$
    \State \textcolor{brown}{//Math Computations.}
    \State $Pipelined\_Shared2Reg\_TensorCoreOps(smem\_r);$
    \State \textcolor{brown}{//barrier: initSharedMem()}
    \State $cp\_async\_wait$\text{<1>}$();$ \text{ } $\_\_syncthreads();$
    \State $ExtractRegister2Shared(smem\_w)$ \Comment{\textcolor{teal}{extract}}
    \State \textcolor{brown}{//barrier: copyGlobal2Shared().}
    \State $cp\_async\_wait$\text{<0>}$();$ \text{ } $\_\_syncthreads();$
\EndFor
\State $results\_Reg2Global(C.data);$

\end{algorithmic}
\end{algorithm}

\subsubsection{Overall Implementation}

Alg.\ref{alg:Kernel_Overview} shows the implementations of the pipelined computation in \SYS{}.
In line 3, the software pipeline will be initialized, preparing the data of $A_{Tile}$ and $B_{Tile}$ on shared memory for the tensor core computations of the first iteration in the main loop.
The following iterations described in Fig.\ref{fig:DataTransfer_Computation_Pipeline} is implemented in lines 7-28.
For each iteration, it issues the instructions for the asynchronous data loading for the next iteration and does the tensor core computation of the current iteration in a double buffer manner.
Specifically, one $A_{Tile}$ for the next iteration will be loaded and extracted from global memory to shared memory (\textit{rst\_smem}, \textit{gmem2reg}, and \textit{extract}), 
and one dense $B_{Tile}$ will be loaded directly from global memory (\textit{ld\_dense}).
The \textit{rst\_smem} stage is in line 17, where each thread issues \textit{cp.async} operation to set buffer A to zeros.
% In line 19, these asynchronous operations are all committed to group-1.
In line 19, \textit{gmem2reg} is accomplished, where sparse encoding is loaded from global memory to the distributed registers.
% No explicit memory barrier is required for \textit{gmem2reg} since register dependencies are ensured and resolved by the GPU hardware.
The \textit{ld\_dense} stage is in line 20, where the data for $B_{Tile}$ is loaded from global memory to shared memory buffer with \textit{cp.async} operations.
% The \textit{cp.async} operations within \textit{ld\_dense} are issued to load the dense $B_{Tile}$ into buffer B in line 21.
% Note that for all \textit{cp.async} operations, we adopt \textit{cp.async.cg} primitive to bypass the access to GPU's L1 cache.
After launching these asynchronous memory operations, tensor core operations are launched in line 23.
Note that we load dense matrices from shared memory to registers using \textit{ldmatrix.sync} instruction and utilize tensor cores for the core computations by explicitly launching \textit{mma.sync} instruction in the function \textit{Pipelined\_Shared2Reg\_TensorCoreOps()}.

The first async-copy barrier in Fig.\ref{fig:DataTransfer_Computation_Pipeline} is in line 25,
guaranteeing that all asynchronous operations launched in line 17 are completed while the operations launched in line 20 can still be in progress. 
The \textit{extract} stage is in line 26, where the data on the distributed registers are extracted to the shared memory buffer of $A_{Tile}$.
Finally, the async-copy and thread-block barriers are called in line 28 to make sure that all threads have completed their work in this iteration.

Different from dense MatMul where the data size to be loaded from global memory can be inferred by the tile sizes, the size of sparse encoding is determined by the number of non-zeros (\textit{nnz}) within $A_{Tile}$, which is unpredictable.
Before loading the sparse encoding for each tile, \SYS{} identifies its start offset and the length in global memory buffers.
Such information is maintained in \textit{TileOffsets} array, which is stored in global memory (more details in Sec.\ref{subsec:sparse-encoding-and-runtime-parsing}).
To avoid instruction stalls caused by long latency global memory access, this metadata should be pre-fetched.
In lines 5-6, the start offset and the size of sparse encoding for the first iteration are pre-fetched.
At the beginning of each iteration, the start offset and the size of the current $A_{Tile}$ are updated using the value in registers pre-fetched in advance.
In lines 11-12, it pre-fetches the start offset and the length for the next iteration.

\subsection{Sparse Encoding and Runtime Parsing}
\label{subsec:sparse-encoding-and-runtime-parsing}

\subsubsection{Tiled-CSL Format}
\label{subsubsec:Tiled-CSL}

\begin{figure}
    \centering
    \includegraphics[width=0.8\columnwidth]{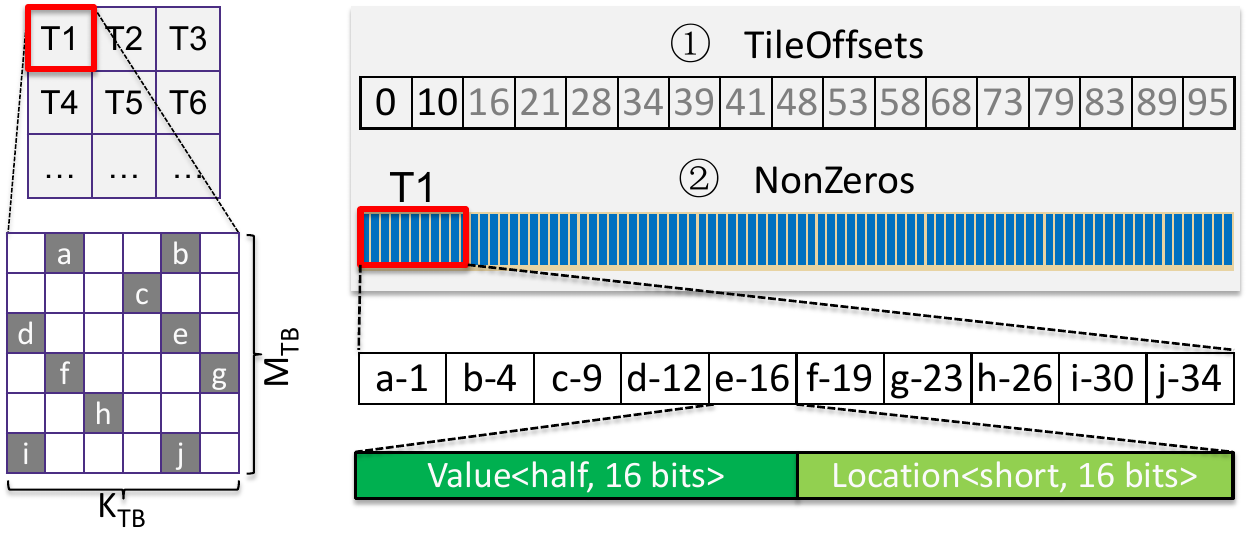}
    \vspace{-4ex}
    \caption{\textit{Tiled-CSL} Format for sparse matrices.}
    \label{fig:Tiled_CSL}
\end{figure}

The sparse encoding format of $A$ matrix is essential for efficient sparse data storage and \textit{Sparse-to-Dense Transformation}.
We propose a tile-by-tile sparse encoding format to support the optimizations in Sec.\ref{subsec:pipeline-design} effectively.
The non-zero elements are organized tile-by-tile, where each tile maintains its non-zero elements accompanied by the sparse index.
As shown in Fig.\ref{fig:Tiled_CSL}, the data of each tile within the sparse matrix are encoded into a small array, and combining all tiles will form the overall array (\textit{NonZeros Array}).
%Note that the tile size of \textit{Tiled-CSL} format is not necessarily the same as that of $A_{Tile}$ in Fig. \ref{fig:LSCD}.
%Assume that the size of \textit{Tiled-CSL} tile is set to $128 \times 64$, $M_{TB}$ in Fig. \ref{fig:LSCD} is set to 128 or 256, and $K_{TB}$ is set to 64.
%Each data tile of $A_{Tile}$ will be represented by one or two \textit{Tiled-CSL} tiles.
% We combine all the arrays together, resulting in the \textit{NonZeros Array} of the entire sparse matrix.
The \textit{TileOffsets Array} maintains the starting offset of each tile in \textit{NonZeros Array}.
The number of non-zero elements of each \textit{Tiled-CSL} tile is the difference between two corresponding elements in \textit{TileOffsets Array}.
% Each tile in the \textit{NonZeros Array} is encoded into the \textit{Tiled-CSL} format, where each non-zero element is stored along with its location within the tile.
For each tile in \textit{NonZeros Array}, each element is stored along with its location within the tile.
As shown in Fig.\ref{fig:Tiled_CSL}, each non-zero weight is in 16-bit half-precision and each location is encoded into a 16-bit short integer.
%The 16-bit short integer is enough to maintain the location within our \textit{Tiled-CSL} tile in shape $128 \times 64$.

\subsubsection{Register to Shared Memory Data Extraction based on Tiled-CSL Format}
\label{section:Runtime-Parsing}

% In this section, we describe the detailed design of \textit{extract} stage.
As described in Alg.\ref{alg:ExtractRegister2Shared}, each thread extracts $nnz\_thread$ non-zero values from its sparse encoding buffer \textit{Reg[]} to the shared memory buffer \textit{A} with a loop.
% For each iteration, one non-zero value is extracted and stored in the dense buffer \textit{A} in shared memory.
The \textit{v()} and \textit{idx()} functions are used to extract the value (high 16 bits) and its location (low 16 bits).
There are some special considerations when using registers as intermediate buffers for sparse encoding.
Different from shared memory and global memory, GPU registers are not addressable, which means we can not access an array of registers using a variable offset.
As a result, forcing an array defined in CUDA into registers requires that, all the indices used to access the array can be determined statically at compile-time.
Otherwise, the array will be stored in global memory instead.
In line 1 in Alg.\ref{alg:ExtractRegister2Shared}, \textit{\#pragma unrol} is used to notify the GPU compiler to fully unroll the main loop, so that all the indices used to access the $Reg[]$ can be determined statically.
Note that adding such compiling directive alone is not enough as a loop with a variable number of iterations can not be fully unrolled.
Thus, we use a constant value $\#REG$ (the upper bound of the number of iterations, typically 32/64 in our implementation) in line 2 instead of using the variable value $nnz\_thread$.
% A breaking logic is used in lines 3-4 to enable early exit.

\begin{algorithm}\footnotesize
\caption{\footnotesize{ExtractRegister2Shared}}
\label{alg:ExtractRegister2Shared}
\begin{algorithmic} [1]
\State \#pragma unroll
\For {\textcolor{blue}{int} $i = 0;$ $i < \#REG;$ $i++$}
    \If{$i \geq nnz\_thread$}
        \State $\textbf{break}$
    \EndIf
    \State $A[ idx(Reg[i]) ] = v( Reg[i] )$
\EndFor
\end{algorithmic}
\end{algorithm}

\iffalse
The registers reserved should be able to store a tile with an arbitrary level of sparsity.
Given that an extra 16-bit will be used to store the index for each 16-bit weight, one register will be used to store one non-zero weight.
If the sparsity within the tile is more than 50\%, this tile will be stored with our CSL format and the total number of registers required is $M_{ThreadBlock}*K_{ThreadBlock}$.
Otherwise, this tile will be stored in the dense format and no \textit{Sparse-to-Dense Transformation} is required.
\fi

\subsubsection{Ahead of Time Sparse Data Reordering}
\label{subsection:Ahead-of-time-Reordering}

There are two types of shared memory access in the computation pipeline for the sparse weight matrix $A$.
The one is shared memory load for tensor core computation in \textit{smem2tc} stage.
The other one is shared memory store in \textit{extract} stage.
% \textit{ExtractRegiter2Shared()} (Alg.\ref{alg:ExtractRegister2Shared}).
It is essential to avoid bank conflict for good performance.
However, the random sparsity makes it challenging to avoid bank conflict of both shared memory load and store.

\begin{figure}
    \centering
    \subfloat[The \textbf{"ldmatrix"} intrinsic loads an 8x8 matrix from shared memory to registers as tensor core inputs. 
    %For efficient matrix reading, each value within the $A_{Tile}$ should be stored in a specific shared memory bank.
    \label{fig:ldmatrix}]{
        \includegraphics[width=0.7\columnwidth]{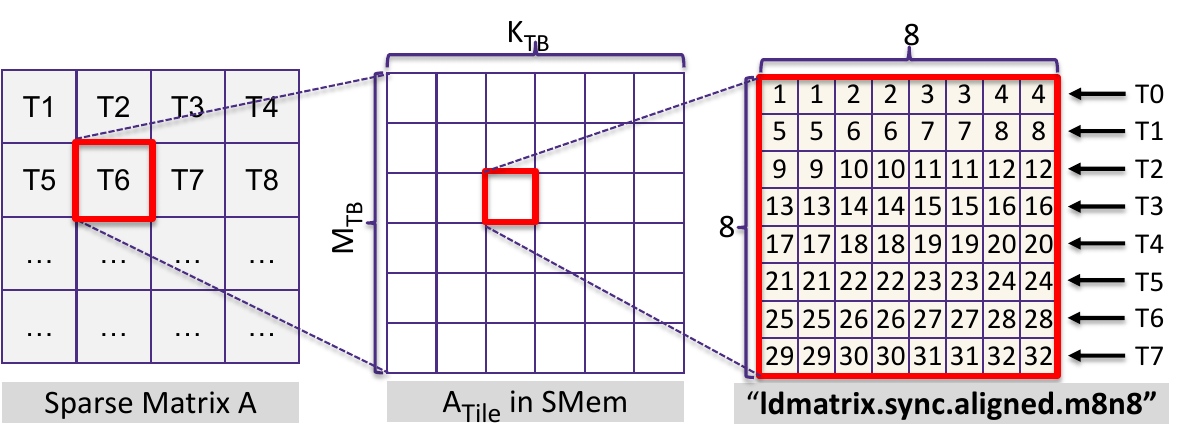}
    }
    \\
    \vspace{-2ex}
    \subfloat[Bank conflicts during \textit{ExtractRegister2Shared}. 
    (WARP size and SMem banks are reduced from 32 to 4 for simplicity.)
    \label{fig:Before-Reorder}]{
        \includegraphics[width=0.9\columnwidth]{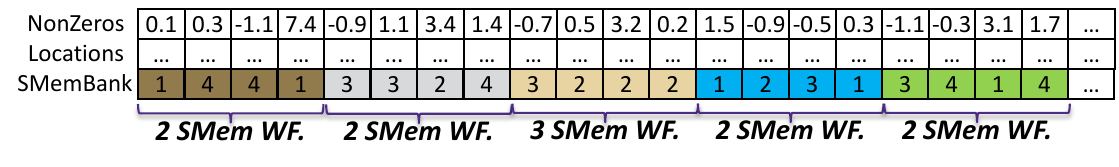}
    }
    \\
    \vspace{-2.5ex}
    \subfloat[
    No conflict after Sparse Data Reorder.
    \label{fig:After-Reorder}]{
        \includegraphics[width=0.9\columnwidth]{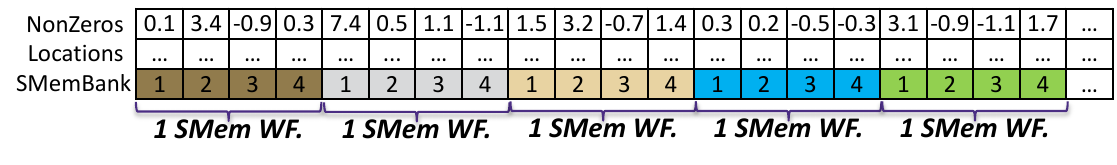}
    }
    \vspace{-2.5ex}
    \caption{Ahead of time sparse data reordering.}
    \label{fig:NonZero-Reorder}
\end{figure}

As for the \textit{smem2tc} stage, it makes use of \textit{ldmatrix} intrinsic for efficient data loading from shared memory to registers for tensor core computation.
Fig.\ref{fig:ldmatrix} shows the behavior of \textit{ldmatrix} where eight threads collectively load an $8\times8$ matrix in FP16 from shared memory.
This $8\times8$ matrix reading can be served by a single shared memory wavefront\footnote{A \textit{wavefront} is the maximum unit of work that can pass through the GPU hardware pipeline per cycle. 
At most 1,024 bits can be loaded per wavefront for shared memory.}
if there is no bank conflict.
%The bank-conflict-free memory load of \textit{ldmatrix} requires that different rows corresponding to the 8 threads in Fig.\ref{fig:ldmatrix} are stored in disjoint memory banks.
The bank-conflict-free memory load of \textit{ldmatrix} requires that all scalars within the $8\times8$ matrix can be read from disjoint memory banks.
Fig.\ref{fig:ldmatrix} shows an example shared memory data layout demonstrating the bank assignment (bank ID ranging from 1 to 32) to achieve bank-conflict-free \textit{ldmatrix}.
Thus, each scalar can be assigned a bank ID according to its location within $A_{Tile}$, given a specific data layout.

However, this requirement will easily cause bank conflict of shared memory store during \textit{extract} stage due to the random position of the sparse elements in matrix $A$.
In other words, we can guarantee the bank-conflict-free load according to the layout requirement of \textit{ldmatrix}, but cannot avoid bank conflict store in this way during \textit{extract} stage.
Fig.\ref{fig:Before-Reorder} gives an example,
where each non-zero (NonZero) value should be stored in a target shared memory bank (SMemBank) according to their relative position within $A_{Tile}$ to meet the requirement of bank-conflict-free \textit{ldmatrix}.
As the distribution of NonZeros is random, the target SMemBank for each NonZero value is also random. 
As a result in Fig.\ref{fig:Before-Reorder}, all CUDA WARPs (the NonZeros with the same color are processed by the same WARP) suffer from bank conflict and lead to multiple shared memory wavefront (SMem WF).
% As shown in Fig.\ref{fig:reduce-bank-conflict}, the NonZeros with the same color are processed by the same WARP at the same time.
% Before reordering, bank conflicts occur unless 32 threads read exactly 32 different banks, which is difficult to guarantee without proper scheduling.

To reduce the bank conflict, we propose the \textit{ahead of time sparse data reordering} approach.
The basic insight is that the bank-conflict-free \textit{ldmatrix} already determines the target bank of each data element,
we can thus reorder the data elements in each \textit{Tiled-CSL} tile so that elements correspond to different banks could be organized into the same WARP for \textit{extract} execution.
Specifically, it iteratively selects the sparse element that corresponds to different memory banks when generating the NonZeros sub-array for each \textit{Tiled-CSL} tile.
Fig.\ref{fig:After-Reorder} shows an ideal case where only one shared memory wavefront is needed to serve one WARP executing \textit{extract} stage after data reordering.
Note that the data reordering is within \textit{Tiled-CSL} tile scope, which only changes the data distribution on global memory but does not change that on shared memory buffers.
In other words, it is a way to change the data placement on global memory to achieve more efficient shared memory access.

\begin{algorithm}\footnotesize
\caption{\footnotesize{Tiled-CSL\_Gen\_AOTSparseDataReordering}}
\label{alg:NonZeroReorder}
\begin{algorithmic} [1]
\State \textbf{Input:} \textit{Matrix A in size $M \times K$;}
\State \textbf{Output1:} \textit{vector< vector<unsigned int> > NonZeros;}
\State \textbf{Output2:} \textit{vecotr<int> TileOffsets;}
\State \textit{vector<unsigned int>} $NZ\_Bucket[32];$
\For {\textcolor{blue}{int} $i = 0;$ $i < M/128;$ $i++$}
    \For {\textcolor{blue}{int} $j = 0;$ $j < K/64;$ $j++$}
        \State \textcolor{brown}{// classifying NonZeros.}
        \State \textcolor{blue}{half} $*TilePTR = A + i*128*K + j*64;$
        \For{\textcolor{blue}{int} $x = 0;$ $x < 128;$ $x++$}
            \For{\textcolor{blue}{int} $y = 0;$ $y < 64;$ $y++$}
                \State $val = TilePTR[x*K+y];$
                \State \textcolor{blue}{short} $loc = location\_in\_SMem(x,y);$
                \State \textcolor{blue}{int} $BankID = (x\%8)*4 + (y\%8)//2;$
                \State $NZ\_Bucket[BankID].push\_back(val, loc);$
            \EndFor
        \EndFor
        \State \textcolor{brown}{// iteratively picking 32 NonZeros as a group.}
        \State \textcolor{blue}{int} $NNZ = count\_NNZ(NZ\_Bucket);$
        \For{\textcolor{blue}{int} $g = 0;$ $g < NNZ/32;$ $g++$}
            \State \textit{vector<unsigned int>} $NZ\_group;$
            \For{\textcolor{blue}{int} $b = 0;$ $b < 32;$ $b++$}
                \State \textcolor{blue}{int} $id = BankID\_Max(NZ\_Bucket);$
                \State $NZ\_Group.push\_back(NZ\_Bucket[id].back());$
                \State $NZ\_Bucket[id].pop\_back();$
            \EndFor 
            \State $NonZeros.push\_back(NZ\_group);$
            \State $TileOffset.push\_back(NNZ);$
        \EndFor
    \EndFor
\EndFor
\end{algorithmic}
\end{algorithm}

Alg.\ref{alg:NonZeroReorder} shows the algorithm to generate \textit{Tiled-CSL} format from the original sparse matrix according to \textit{ahead of time sparse data reordering} approach.
The input is the sparse matrix A with $M$ rows and $K$ columns in dense format,
where some elements are already set to 0 through model pruning.
The outputs are \textit{NonZeros} and \textit{TileOffsets}, the key components of \textit{Tiled-CSL} format.
\textit{NonZeros} are split into groups each of which contains 32 non-zeros in line 2. 
Each group of non-zeros will be written to shared memory during \textit{extract} stage within one shared memory request.
At lines 7-24, the \textit{Tiled-CSL} format of one tile ($128 \times 64$) will be generated.
At lines 11-14, each non-zero within matrix A will be encoded into a 32-bit word containing $(val, loc)$.
Besides, non-zeros are distributed to 32 different buckets ($NZ\_Bucket[32]$ at line 4) according to their target shared memory bank ID ranging from 0 to 31 as calculated in line 13. 
In line 16, the total number of non-zeros (NNZ) is counted.
At line 19-22, one group of non-zeros are formed by iteratively picking non-zeros from \textit{NZ\_Bucket[id]} where $NZ\_Bucket[id]$ is the bucket with the most non-zeros not processed at that time.
% Greedily picking non-zeros from the bucket with the most number of unassigned non-zeros ensures two things:
% (1) Keeping the overhead of unavoidable bank conflicts small. 
% At the very beginning, there will be bank conflicts within the formed non-zero groups, which is unavoidable as the numbers of non-zero for different banks may be different.
% Thus, we can not always make the perfect schedule shown in Fig.\ref{fig:After-Reorder}. 
% (2) Forming non-zero groups with no bank conflict for the rest of the groups.

\section{Implementation} 
\label{section:System-Integration}

% We implement \SYS{} with 1.8 K lines of C++/Python code (0.8 K lines of C++ for the CUDA kernels and 1 K for other infrastructures). 

% \paragraph{Kernel Integration}
We provide a set of C++ APIs for high-performance \SYS{} kernel.
We integrate \SYS{} kernel into FasterTransformer~\cite{Faster-Transformer}, enabling high-efficiency distributed inference with sparsified weight matrices.
Specifically, we extended its corresponding class definition (i.e. \textit{DenseWeight} class) to support the \textit{Tiled-CSL} format.
Besides, we extended its library wrapper (i.e., \textit{cuBlasMMWrapper} class) to support calling either the dense MatMul library or \SYS{} SpMM kernel according to the given data format.
% Note that we use FT for inference using its C++ interfaces in this paper.
\SYS{} can also be easily integrated into other deep learning frameworks through library calls with \SYS{} API.
We also provide a weight reformatting tool to generate sparse matrices in Tiled-CSL format given the pre-trained dense PyTorch model.

In our implementation, the size of $M_{TB}$ in Fig. \ref{fig:LSCD} is 128 or 256, $K_{TB}$ is 64, and $N_{TB}$ is 8/16/32/32 when the $N$ dimension of MatMul (inference batch size) is 8/16/32/64.
For larger $N$ dimensions, the $N_{TB}$ is 64.
The thread block size is 128.
% We also make sure that each GPU SM can accommodate two thread blocks simultaneously when planning the register/SharedMem usage.
These configurations work well for the workloads we evaluated in Sec.\ref{sec:evaluation}.
The configurations can be reconfigured easily for other workloads.
The configuration tuning is not in the research scope of this paper.

\section{Evaluation}
\label{sec:evaluation}

\begin{figure*}[hbt!]
    \centering
    \includegraphics[width=\linewidth]{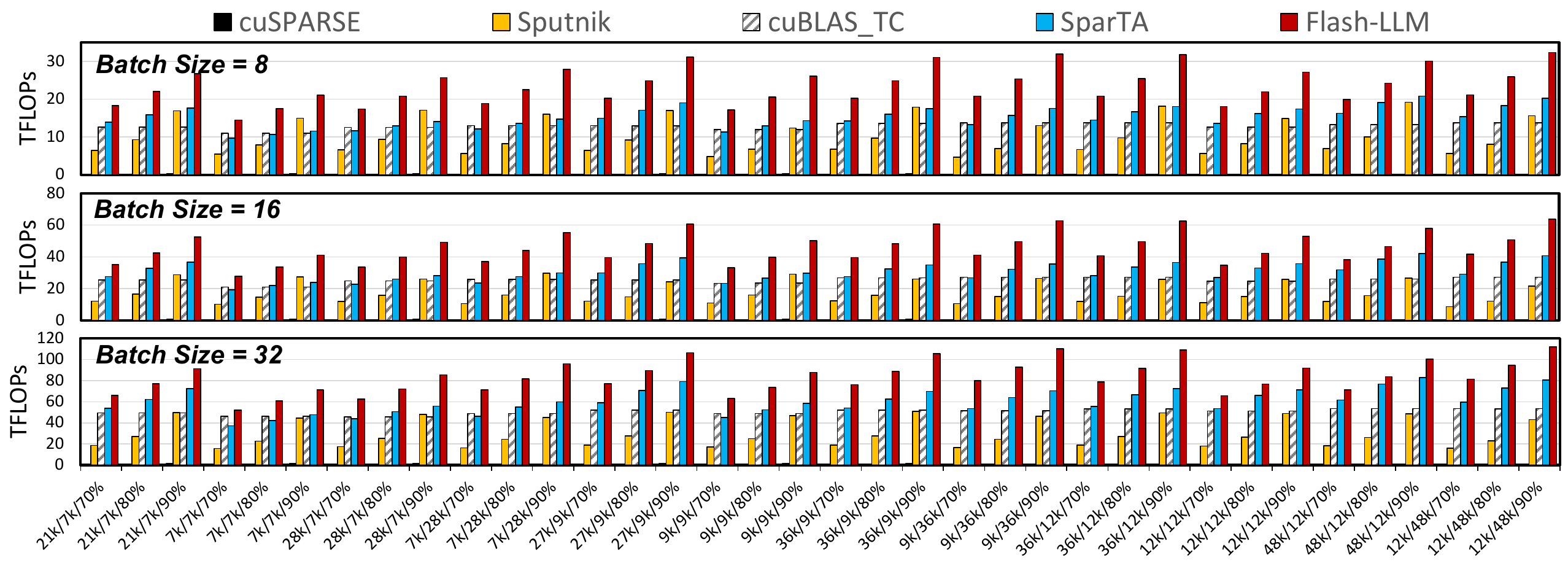}
    % \vspace{-6ex}
    \caption{Kernel Benchmarking (M/K/Sparsity; weight matrix: M $\times$ K).}
    \label{fig:KernelBenchmark}
\end{figure*}

We evaluate the performance of \SYS{} on two levels: kernel-level benchmarking and model-level evaluation.
The evaluation is conducted on the NVIDIA A100-SMX8-80GB platform (128-core Intel Xeon Platinum 8369B CPU @2.90GHz, 8 NVIDIA A100 GPU @80GB), with Ubuntu 18.04 and CUDA 11.8.
We enable auto-mixed precision (AMP) for all evaluations.
% \cb{
% \marginpar{\cb{R2.W2}}
We mainly do experiments on NVIDIA A100 GPUs.
However, the method we proposed is also a good reference for the kernel design of TPU and Intel CPUs that are equipped with customized hardware for matrix multiplication.
% }
% For kernel evaluation, we compare \SYS{} with the state-of-the-art SpMM kernels for unstructured sparsity.
% For model evaluation, we compare the throughput of \SYS{} with FasterTransformer (FT)~\cite{Faster-Transformer} and DeepSpeed (DS)~\cite{DeepSpeed-Inference}.

\subsection{Kernel Performance}

\parahead{Workloads, baselines, and settings.}
We evaluate \SYS{} on MatMuls under different shapes, coming from the four MatMuls described in Sec.\ref{sec:GenerativeModelInference} within OPT-30B, OPT-66B, and OPT-175B~\cite{OPT-Models} given four different batch sizes (8, 16, 32, and 64).
% For each model, the performances of four types MatMuls are evaluated, including the QKV projection, the attention output project, and the two fully-connected layers.
% , as described in Fig.\ref{fig:Decoder-Layer-Architecture}.
%MalMuls related to \textit{self\_attn\_qkv\_proj.weight}, \textit{self\_attn\_out\_proj.weight}, \textit{fc1.weight} and \textit{fc2.weight}.
% We evaluate the MatMul latency with four different batch sizes (8, 16, 32, and 64).
% Thus, there are 3 models $\times$ 4 MatMul per model $\times$ 4 batch sizes = 48 MatMul shapes.
For each MatMul shape, we evaluate the kernel latency under 70\%, 80\%, and 90\% of random sparsity in the weight matrices.
% In total, 144 test cases are used.
The baselines we compare include cuSPARSE~\cite{cuSPARSE}, Sputnik (commit: 46e380c)~\cite{Sputnik-Github, Sputnik}, SparTA (commit: 1f61a36)~\cite{SparTA-Github, SparTA}, and cuBLAS~\cite{cuBLAS}.
CuSPARSE is part of the CUDA Toolkit for handling sparse matrices.
Sputnik is a library of sparse linear algebra kernels for deep learning,
which achieves state-of-the-art SpMM performance based on SIMT cores.
SparTA supports unstructured sparsity based on 2:4 structured sparsity on tensor core~\cite{SparseTensorCore-NVIDIA}.
% SparTA proposes to support unstructured sparsity leveraging both 2:4 sparsity kernels and Sputnik kernels.
% It utilizes Ampere GPU's sparse TensorCore\cite{SparseTensorCore-NVIDIA} to exploit 2:4 sparsity by using the API provided by cuSPARSELt~\cite{cuSPARSELt} library.
As SparTA only supports FP32 precision, we extended SparTA to support input matrices in FP16.
CuBLAS targets dense MatMul rather than SpMM.
We include it as a baseline here to show the practical performance gains/loss compared to the basic dense implementations of LLM inference.
% When benchmarking the performance of cuBLAS and \SYS{}, the dense matrix B and output matrix C are stored in column-major as it is the default mode for cuBLAS.
% When evaluating cuSPARSE/Sputnik/SparTA kernels, row-major is used since Sputnik/SparTA only supports row-major while cuSPARSE performs better under row-major. 

\parahead{Results.}
Fig.\ref{fig:KernelBenchmark} shows the kernel performance (TFLOPs) of \SYS{} and the baselines.
% We use Tera Float-point Operations per second (TFLOPs) as the main metric for comparison.
The throughput can be calculated by $2 \times M \times K \times N / kernel\_latency$.
As shown in Fig.\ref{fig:KernelBenchmark}, \SYS{} performs the best constantly compared to the baselines.
On average, \SYS{} outperforms Sputnik/SparTA by $3.6\times$/$1.4\times$, $3.0\times$/$1.4\times$, and $2.0\times$/$1.6\times$ under 70\%, 80\%, and 90\% sparsity respectively.
Besides, \SYS{} can also outperform the state-of-the-art dense kernels cuBLAS with tensor core enabled by $1.4\times$, $1.7\times$, and $2.1\times$.
CuSPARSE shows poor performance under such moderate-level sparsity, as it is designed for matrices with >95\% sparsity\cite{cuSPARSE}.
% Sputnik is designed for pruned deep learning models, thus it works well under this range of sparsity.
% Aligned with the claim in their paper\cite{Sputnik}, Sputnik outperforms SIMT-core-based dense kernels.
As for Sputnik, it is very challenging to outperform cuBLAS kernels with tensor core enabled.
According to the benchmark results, Sputnik can outperform cuBLAS\_TC only when the sparsity is >90\% when the N dimension is 8, 16, or 32.
%For larger N dimensions, which means larger inference batch sizes, Sputnik shows much slower throughput compared to cuBLAS\_TC, as the limited opportunities for skipping computations under moderate-level sparsity can not make up the huge performance gap between SIMT cores (Sputnik) and tensor cores (cuBLAS\_TC).
As for SparTA, as described in Sec.\ref{section:Demand-of-Unstructured-Sparsity}, it leverages sparse tensor core~\cite{SparseTensorCore-NVIDIA} for the major part of the computations, which can not effectively exploit the sparsity available as sparse tensor cores only support 50\% sparsity (2:4 sparsity).
If the sparsity available is higher than 50\%, SparTA has to pad zeros to the sparse matrix resulting in redundant global memory access during runtime.
Besides, for the non-zeros that can not meet the 2:4 requirement, another SIMT core based kernel is launched for the corresponding computations, resulting in extra overhead.
Therefore, \SYS{} outperforms SparTA in our evaluations.

\begin{figure}
    \centering
    \subfloat[Batch size = 16 \label{fig:Profiling-BS16}]{
        \includegraphics[width=0.99\columnwidth]{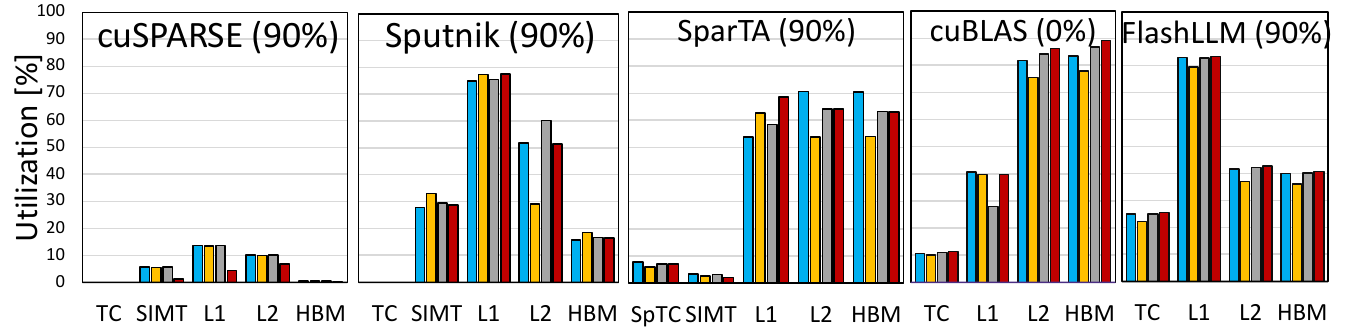}
    }
    \\
    % \vspace{-3ex}
    \subfloat[Batch size = 32 \label{fig:Profiling-BS32}]{
        \includegraphics[width=0.99\columnwidth]{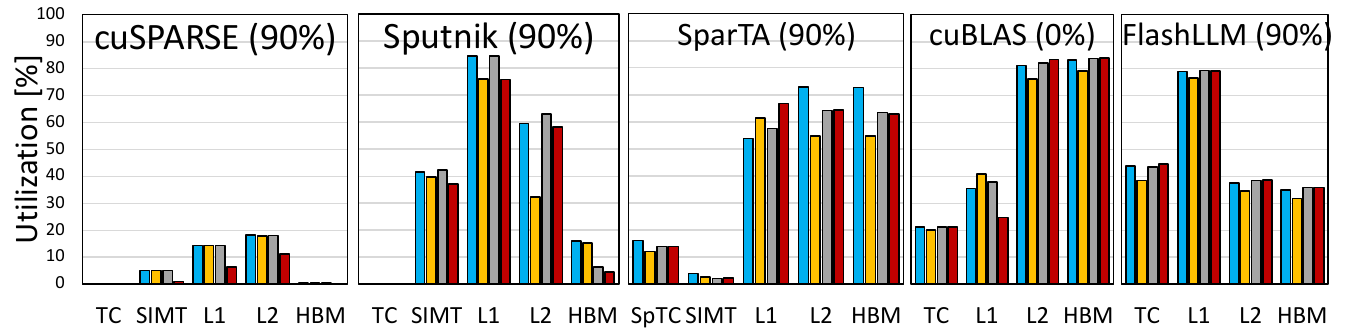}
    }
    % \vspace{-3ex}
    \caption{ 
    Kernel utilization breakdown with four MatMul shapes (indicated with different colors) from OPT-66B. }
    \label{fig:KernelProfiling}
\end{figure}

\subsection{Kernel Analysis}

% \marginnote{
% \footnotesize
% \cb{R1.D5}
% }[-2.9in]
\parahead{Optimized GPU Utilization.}
Fig.\ref{fig:KernelProfiling} 
\footnote{For cuBLAS\cite{cuBLAS}, the number of tensor core instructions (HMMA) executed will not change as the N dimension varies (e.g. 8, 16, 32, 64).
It means that there are redundant tensor core operations launched in cuBLAS kernels for N=8, 16, and 32. 
These redundant HMMA operations are excluded properly in our evaluations}.
shows the utilization of GPU hardware units including tensor cores (TC), combined L1 and shared memory (L1), L2 cache (L2), and GPU global memory (HBM) during \SYS{} kernel execution.
All the data is collected by the NSight Compute profiler~\cite{NsightCompute}.
% \marginpar{\cb{R1.D5}}
% \cb{
We present the profiling results of N = 16 and N = 32 under 90\% sparsity.
We also include cuBLAS here as the dense baseline (sparsity = 0\%).
cuSPARSE and Sputnik are SIMT-based designs where tensor cores are not used at all.
Although Sputnik achieves good SIMT core utilization (29.8\% at BS=16, 40.1\% at BS=32), it achieves much slower performance than other tensor-core-based kernels as SIMT cores show much lower peak computational throughput than tensor cores.
SparTA utilizes the tensor core but with lower utilization than cuBLAS.
% }
For cuBLAS, the bandwidth of the L2 cache and GPU DRAM is exhausted while the tensor cores only reach 10.7\% and 21.0\% of its peak performance on average when the N dimension is 16 and 32 respectively.
In \SYS{}, the sparse matrices are in \textit{Tiled-CSL} format (Sec.\ref{sec:sparse-load-dense-compute}) with reduced size in bytes.
% \cb{
% \marginpar{\cb{R1.D5}}
The global memory bandwidth is no longer the bottleneck with the \textit{Load-as-Sparse and Compute-as-Dense} method.
On average, tensor core utilization is improved to 24.4\% and 42.6\%.
As tensor core utilization is improved, it consumes higher bandwidth of shared memory.
% }
In addition to that, the \textit{extract} in Fig.\ref{fig:DataTransfer_Computation_Pipeline} will cause extra shared memory stores.
Thus, L1/shared-memory bandwidth is exhausted by \SYS{}, which prohibits further performance improvements.
Note that the \textit{ahead of time sparse data reordering} (Sec.\ref{subsection:Ahead-of-time-Reordering}) is designed to increase the shared memory access efficiency, which helps to mitigate the bandwidth bottleneck of shared memory.
How to further reduce the pressure on shared memory could be future work.

\begin{figure}
    \centering
    \subfloat[Batch size = 16 \label{fig:Latency-BS16}]{
        \includegraphics[width=0.48\columnwidth]{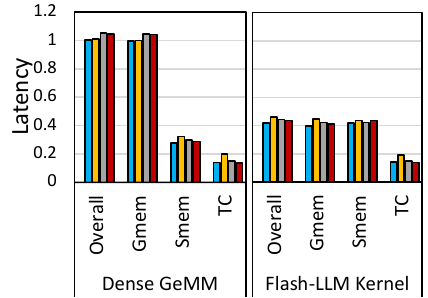}
    }
    \subfloat[Batch size = 32 \label{fig:Latency-BS32}]{
        \includegraphics[width=0.48\columnwidth]{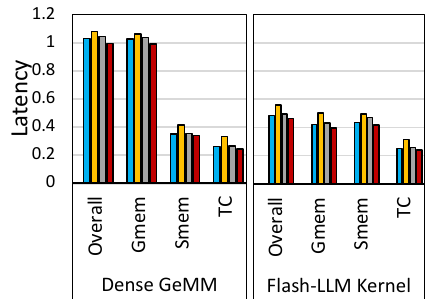}
    }
    % \vspace{-3ex}
    \caption{ Latency breakdown of Dense and \SYS{} Kernels (normalized to cuBLAS\cite{cuBLAS} kernel latency).
    }
    \label{fig:Latency}
\end{figure}

\parahead{Balanced pipeline for memory/tensor core operations.}
% \cb{
% \marginpar{\cb{R1.D3}}
\SYS{} kernel contains three major types of operations: global memory access (Gmem), shared memory access (Smem), and tensor core operations (TC).
Ideally, these three types of operations should be overlapped and conducted in parallel for maximum GPU hardware utilization.
As shown in Fig.\ref{fig:Latency}, We measure the latency of each type of operation separately by erasing other computations in the source code.
% By manually excluding the unrelated operations in the kernel source code, we succeed in measuring the latency of each type of operation separately.
We also re-implement the dense GeMM kernel based on the design of NVIDIA cutlass \cite{cutlass}, which achieves similar performance compared to cuBLAS.
% Thus, we are also able to change the source code of Dense GeMM kernels (note that cuBLAS is a closed source) and measure their latency breakdown for comparison.
% As shown in Fig.\ref{fig:Latency}, our re-implemented GeMM kernel achieves similar performance compared to cuBLAS.
Due to the Buckets effect, the overall kernel time is mainly determined by the global memory operations which require the longest execution time.
With \textit{Load-as-Sparse and Compute-as-Dense} method, the latency of Gmem operations is significantly reduced, leading to overall performance improvements.
Although the Smem latency is increased due to the extra shared memory access required by the \textit{extract} in Fig.\ref{fig:DataTransfer_Computation_Pipeline}, it does not prevent \SYS{} kernel from achieving better performance than cuBLAS.
It is a good trade-off between global memory and shared memory utilization.
What's more, \SYS{} shows similar latency with dense GeMM kernel in terms of tensor core operations as we do not skip any computations here.
However, it is clear that tensor core operations are not the bottleneck for the overall kernel performance.
% }

%To mitigate this problem, we try to reduce memory access of shared memory in this paper as much as possible.
%For example, we choose to store sparse encoding loaded from global memory in registers instead of storing it in shared memory as described in section \ref{section:Overview} to avoid the turn-around shared memory access.
%In addition to that, we propose \textit{Ahead-of-time Reordering of Non-Zeros} in section. \ref{subsection:Ahead-of-time-Reordering} to reduce shared memory bank-conflict.
%cuBLAS contains redundant TC instructions, which results in false high utilization.

\begin{figure}
    \centering
    \includegraphics[width=0.8\columnwidth]{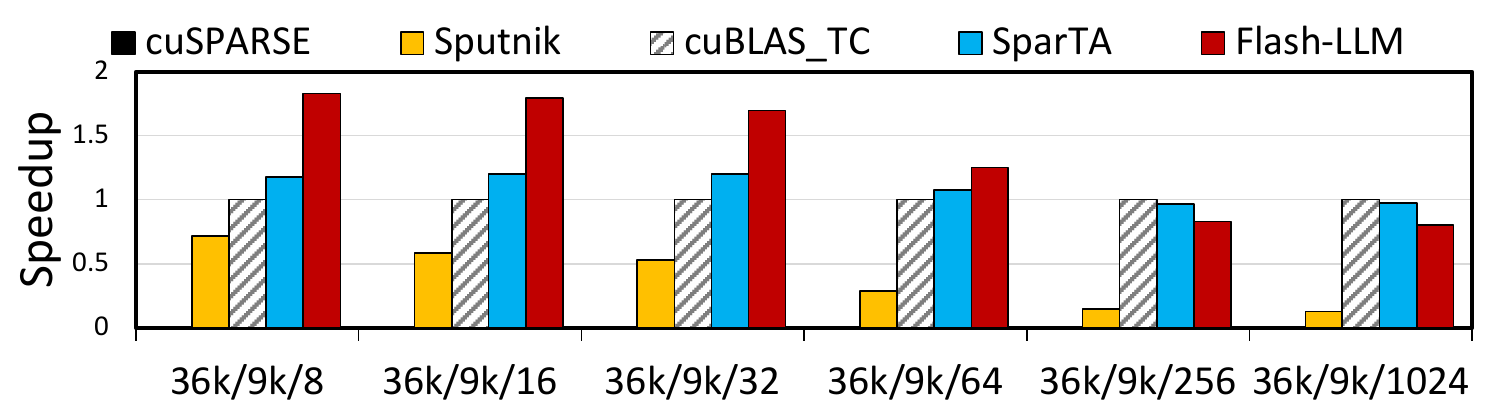}
    % \vspace{-3ex}
    \caption{ Kernel speedups over cuBLAS \cite{cuBLAS} GeMM kernel with different shapes (M/K/N, sparsity=80\%).
    }
    \label{fig:NonSkinny}
\end{figure}

% \marginnote{
% \footnotesize
% \cb{R1.D3}
% }[-7.7in]
\parahead{Performance on more MatMul shapes.}
% \cb{
% \marginpar{\cb{R2.D1\\R2.W1}}
As discussed in Sec.\ref{subsubsec:skinny-matmul-bottleneck}, we mainly want to mitigate the inefficiency caused by Skinny MatMuls in common LLM inference.
For more comprehensive understanding evaluations, we provide kernel performance on more shapes even when the shape is not common for LLM inferences.
As shown in Fig.\ref{fig:NonSkinny}, 
% \marginpar{\cb{R1.D3}}
\SYS{} becomes slower than cuBLAS if the N dimension is larger than 256, noting the memory footprint of \SYS{} is still smaller than dense counterparts.
The reason behind this is twofold.
First, the inefficiency of cuBLAS is no longer significant as the N dimension is large enough, which makes cuBLAS more performant.
Second, \SYS{} has more complicated kernel designs and extra shared memory access, which slows it down a little bit.
We also notice that Sputnik becomes much slower than other tensor-core-based designs as the N dimension increases.
Even though SIMT-core-based solutions such as Sputnik can skip computations exploiting sparsity, such computational savings still can not make up for the huge performance gap between SIMT cores and tensor cores.
% }

%\parahead{Discussion.}
%Discussion on Other Shapes (N very large)
%Our SpMM 20\% slower than cuBLAS\_TC when N=8192, Sputnik $10\times$ %slower than cuBLAS\_TC when N=8192.
%8192 = BS64 * InputSeqLen128

%Discussion on Other Sparsity (Sparsity >= 99\%)
%cuSPARSE and Sputnik show advantages

\subsection{End-to-End Model Evaluation}

% \marginnote{
% % \footnotesize
% \cb{R2.D1}
% }[-4.7in]

\parahead{Baselines.}
We include NVIDIA FasterTransformer (FT) (git commit: 9770d30)~\cite{Faster-Transformer} and DeepSpeed (DS)~\cite{DeepSpeed-Inference} as baselines, the state-of-the-art inference frameworks supporting model parallelism~\cite{Megatron-LM} to fit large models that would otherwise not fit in GPU memory.

\begin{table}\footnotesize
  \centering
  \begin{tabular}{|c|c|c|c|c|c|}
    \hline
    \textbf{Batch Size}      & 8 & 16 & 32  & 64 & 128\\
    \hline
    \textbf{Ours-1GPU}       & 34.7 & 40.8  & 52.9   & 77.1 & OOM\\
    \hline
    \textbf{FT-1GPU}        & 62.7 & 68.7  & OOM    & OOM & OOM\\
    \hline
    \textbf{DS-1GPU}        & 64.1 & 70.9  & OOM    &  OOM & OOM\\
    \hline
    \textbf{FT-2GPU}        & 65.6 & 71.7  & 83.9    & 108.3 & OOM\\
    \hline
    \textbf{DS-2GPU}        & 68.3 & 75.7  & 90.6  & 120.4 & OOM \\
    \hline
  \end{tabular}
  \caption{Peak GPU memory usage (GB).}
  \label{table:MemoryUsage}
\end{table}

\parahead{Workloads.}
We benchmark the end-to-end inference latency of 
the OPT models~\cite{OPT-Models}, including OPT-30B, OPT-66B, and OPT-175B. 
% \cb{
% \marginpar{\cb{R3.D1}} 
In order to accommodate only the model parameters in dense format, 
60/132/350GB GPU memory is required for OPT-30B/66B/175B.
Note that SOTA GPUs only have 80GB of memory each, at least 1/2/8 GPUs are required 
\footnote{The number of GPUs must be a power of 2 for model parallelism\cite{ModelParallel}.} 
for the inference of OPT-30B/66B/175B.
Besides, extra GPU memory is required to store the KV-Cache (refer to Fig. \ref{fig:kv-cache}, with its size positively related to inference batch size) during runtime.
According to Table. \ref{table:MemoryUsage}, the existing inference frameworks (FT-1GPU, DS-1GPU) can easily run out of GPU memory during the inference of OPT-30B for batch sizes larger than 16 if storing the model parameters in dense format.
% }
For all experiments, the input/prompt sequence length is 64 and the output/generated sequence length is 512.

\parahead{Metric.}
We propose the metric \textit{tokens per GPU-second} to indicate the normalized inference throughput with the consideration of both execution time and hardware cost (i.e., the number of GPUs used).
% In the following sections, we may compare the performance of two different inference systems where different numbers of GPUs are used.
% Thus, we proposed to use the normalized generation throughput (in \textit{tokens per GPU-second}) as the primary metric for throughput comparisons.
It is calculated with the following equation:
\begin{equation}
    Performance = \frac{N_{token}}{\sum_{i=1}^{N_{GPU}} T_i}
\end{equation}
$N_{token}$ means the number of tokens generated, whereas $N_{GPU}$ and $T_i$ mean the GPU number and the time spent on the i'th GPU for execution.
We use this metric to evaluate the system's performance.
% \cb{
% \marginpar{\cb{R3.D4}}
Note that for real-time inference serving, the inference throughput is the higher the better once the inference latency is less than a specific threshold.
% }

\subsubsection{Case Study: OPT-30B}

\parahead{Model Pruning.}
To show that pruned model has comparable performance with the original model, we evaluate the accuracy of the pruned model with OPT-30B~\cite{OPT-Models} and GPT-NEOX-20B~\cite{GPT-NEOX, GPT-NEOX-HF} on Recognizing Textual Entailment task in SuperGLUE~\cite{SuperGlue}.
To achieve better accuracy, we adopt the popular pruning method Taylor Pruning~\cite{TaylorPruning} to prune these models and keep the front quarter and the last quarter feedforward input layers dense.
Based on that, we achieve 80\% sparsity on OPT-30B and GPT-NEOX-20B with only 1.44\% and 0.72\% accuracy decrease, respectively.
Specifically, accuracy decreases from 85.55\% to 84.11\% on OPT-30B, and from 83.03\% to 82.31\% on GPT-NEOX-20B.

\begin{figure}
\centering
    \subfloat[Single GPU solutions. \label{fig:OPT-30B-InferenceThroughput-1GPU}]{
        \includegraphics[width=0.45\columnwidth]{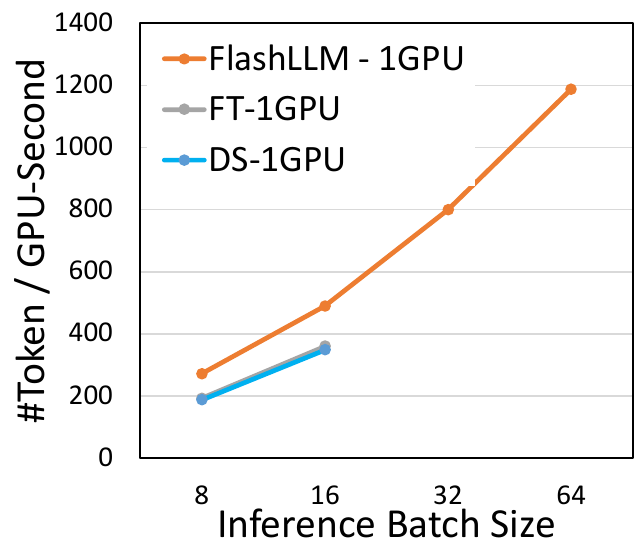}
    }
    \subfloat[Compared to 2-GPU solutions. \label{fig:OPT-30B-InferenceThroughput-2GPU}]{
        \includegraphics[width=0.45\columnwidth]{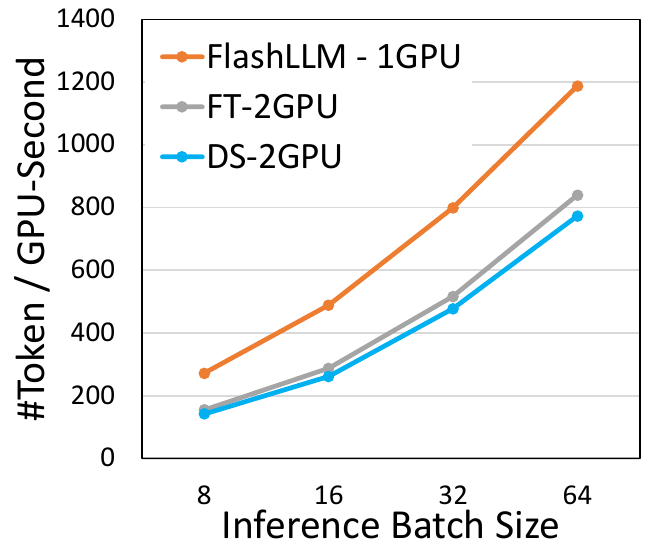}
    }
    \vspace{-1ex}
    \caption{OPT-30B Inference Throughput.}
    \label{fig:OPT-30B-InferenceThroughput}
\end{figure}

\parahead{Results.}
As shown in Fig.\ref{fig:OPT-30B-InferenceThroughput-1GPU}, \SYS{} achieves $3.4\times$ and $3.3\times$ higher performance than DeepSpeed (DS) and FasterTransformer(FT) with a single GPU.
DS and FT can at most achieve 348 and 359 \textit{tokens per GPU-second} with a single A100 GPU.
If further increase the inference batch size, DS/FT will run out of memory as inference tasks with larger batch sizes need more GPU memory to store the cached-KV and activations.
In contrast, \SYS{} achieves up to 1187 \textit{tokens per GPU-second} on batch size 64.
It is because the memory used for storing model weights is reduced with the \textit{Tiled-CSL} format, and thus more Cached-KV and activations can be accommodated.
%We also compare the performance of \SYS{} to DS and FT with model parallelism enabled to make DS and FT support batch size 64, with two-way \textit{model-parallelism}\cite{Megatron-LM} to store the model weights collaboratively using two GPUs.
We also compare the performance of \SYS{} to DS and FT with two-way \textit{model-parallelism}\cite{Megatron-LM}, with which DS and FT can support batch size 64.
% In this way, DS and FT can conduct inference with batch size 64 for this model.
As shown in Fig.\ref{fig:OPT-30B-InferenceThroughput-2GPU}, FT and DS achieve similar performance in terms of \textit{tokens per GPU-second}.
Compared to DS/FT, \SYS{} achieves $1.91\times$/$1.75\times$, $1.87\times$/$1.70\times$, $1.67\times$/$1.55\times$, and $1.54\times$/$1.41\times$ higher performance at batch sizes 8, 16, 32, and 64 respectively.
The detailed GPU memory usage of \SYS{}, FT, and DS are shown in Table.\ref{table:MemoryUsage}.

% \marginnote{
% % \footnotesize
% \cb{R3.M}
% }[-6.0in]

\begin{figure}
\centering
    \subfloat[OPT-30B.
    \label{fig:Breakdown-Of-OPT-30B}]
    {
        \includegraphics[width=0.48\columnwidth]{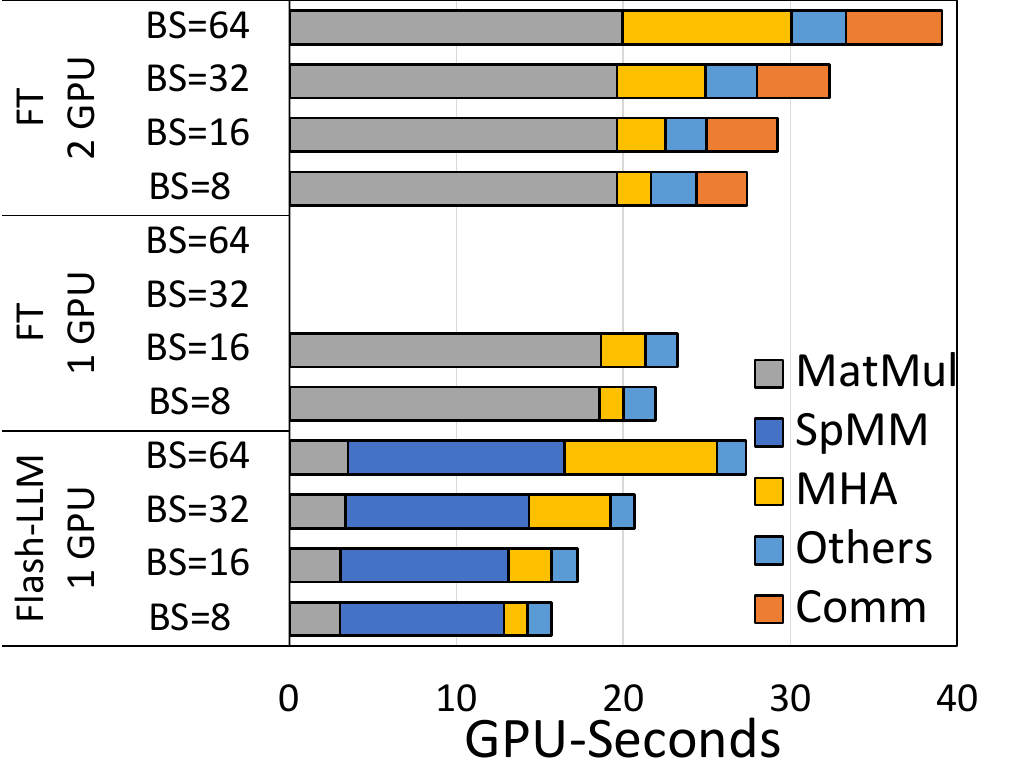}
    }
    \subfloat[OPT-66B.
    \label{fig:Breakdown-Of-OPT-66B}]
    {
        \includegraphics[width=0.48\columnwidth]{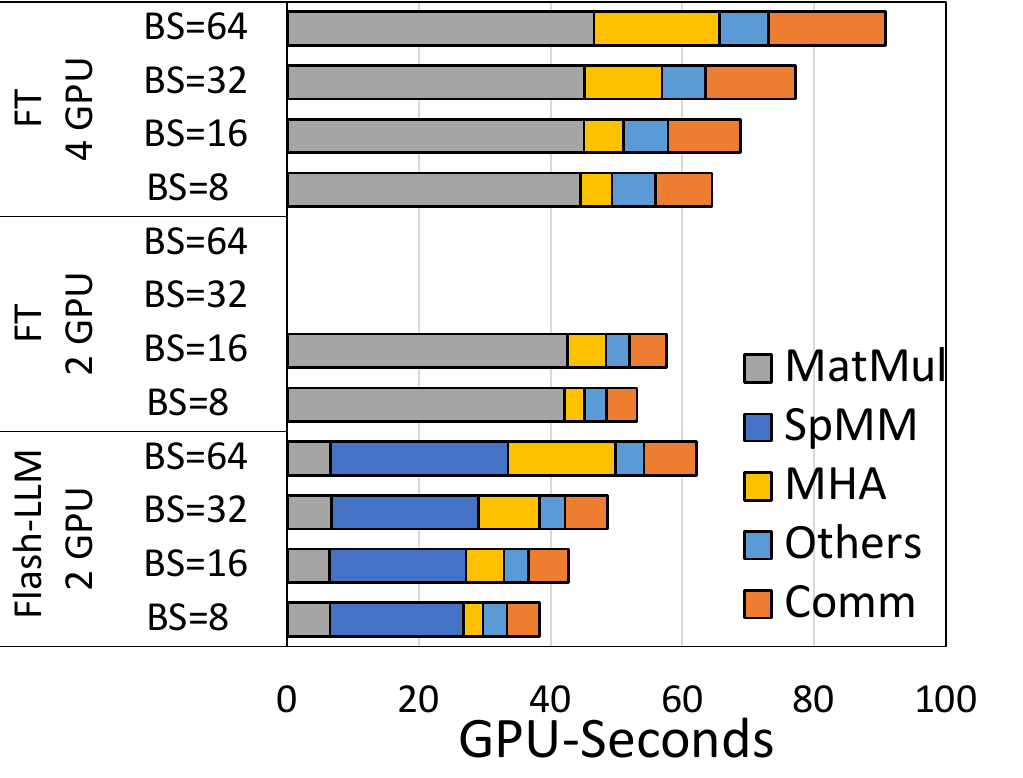}
    }
    % \vspace{-3ex}
    \caption{Inference Time Breakdown. (MHA: multi-head attention, Comm: cross-GPU communications)}
    \label{fig:Breakdown-30B-66B}
\end{figure}

\parahead{Breakdown.}
\label{section:End2EndBreakDown-30B}
In order to figure out why \SYS{} can achieve better performance, we conduct the time breakdown of end-to-end inference shown in Fig.\ref{fig:Breakdown-Of-OPT-30B}.
Note that we conduct all the end-to-end breakdowns in this paper leveraging the NSight System~\cite{NsightSystem}.
Compared to FT-2GPU (FasterTransformer with 2 GPU used), \SYS{} with 1 GPU can achieve lower normalized inference latency \footnote{To also take inference cost into consideration and compare the inference efficiency with different system configs (e.g. different numbers of GPUs may be used), we sum the execution time on all used GPUs as the normalized latency.} mainly because of 1) the more efficient MatMul and 2) the elimination of cross-GPU communication overhead.
%Firstly, the MatMul of \SYS{} is more efficient than cuBLAS.
%Specifically, \SYS{} reduced the MatMul time from 19.6/19.6/19.6/20.0 seconds to 12.8/13.1/14.3/16.5 seconds at batch size 8/16/32/64.
% Note that some weight matrices are still kept in dense to preserve model accuracy, \SYS{} still contains some dense MatMul kernels.
%Besides, we eliminated the cross-GPU communications in FT-2GPU, which cost 3.0/4.2/4.3/5.8 seconds.

%Put them all together, \SYS{} costs 15.7/17.2/20.7/27.3 seconds compared to FT-2GPU which costs 27.3/29.2/32.3/39.1 seconds at batch size 8/16/32/64.
%For instance, FT spends 18.6/18.7 seconds for MatMul computations at batch size 8/16, while our SpMM kernels are used to replace the dense kernels which only cost around 9.8/10.0 seconds.

%\begin{figure}
%    \centering
%    \includegraphics[scale=0.3]{./Figures/AccumulatedKernelTime-OPT30B.pdf}
%    \caption{Breakdown of OPT-30B.}
%    \label{fig:Breakdown-Of-OPT-30B}
%\end{figure}

\subsubsection{Case Study: OPT-66B Model}

\begin{figure}
\centering
    \subfloat[Two GPU solutions. \label{fig:OPT-66B-InferenceThroughput-2GPU}]{
        \includegraphics[width=0.45\columnwidth]{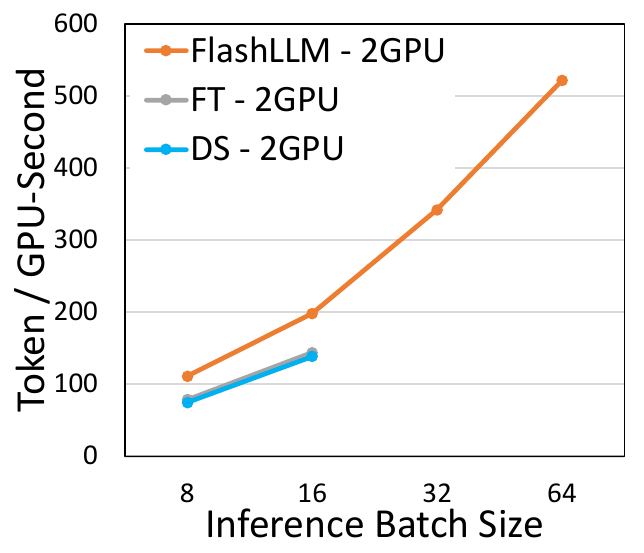}
    }
    \subfloat[Compared to 4-GPU solutions. \label{fig:OPT-66B-InferenceThroughput-4GPU}]{
        \includegraphics[width=0.45\columnwidth]{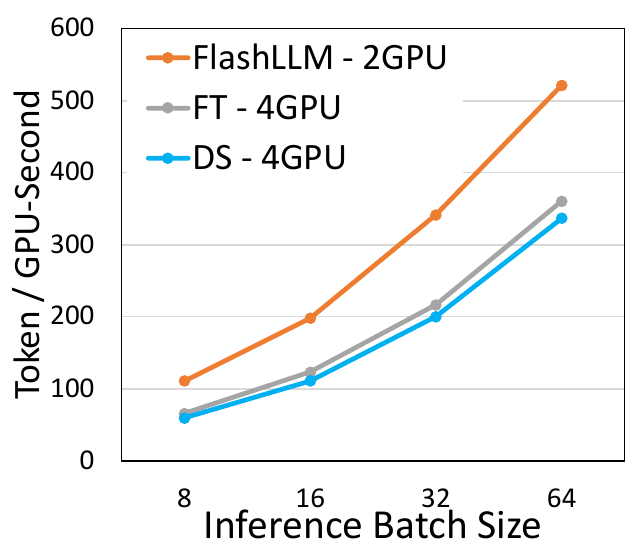}
    }
    % \vspace{-3ex}
    \caption{OPT-66B Inference Throughput.}
    \label{fig:OPT-66B-InferenceThroughput}
\end{figure}

\parahead{Result}
As shown in Fig.\ref{fig:OPT-66B-InferenceThroughput-2GPU}, \SYS{} achieves $3.8\times$ and $3.6\times$ higher token generation throughput than DS and FT with two GPUs.
DS and FT can at most achieve 139 and 144 \textit{tokens per GPU-second} with batch size 16 as they will run out of memory if further increasing the batch size.
In contrast, \SYS{} achieves up to 522 \textit{tokens per GPU-second} at batch size 64.
We also compare the performance of \SYS{} to DS-4GPU/FT-4GPU where \SYS{} still uses two GPUs, while DS-4GPU/FT-4GPU uses four GPUs to enable bigger batch sizes for the baselines.
Compared to DS-4GPU/FT-4GPU, \SYS{} achieves $1.85\times$/$1.68\times$, $1.78\times$/$1.61\times$, $1.7\times$/$1.58\times$, and $1.55\times$/$1.45\times$ higher performance of \textit{tokens per GPU-second} at batch sizes 8, 16, 32, and 64 respectively.

% \marginnote{
% % \footnotesize
% \cb{R3.M}
% }[-5.7in]

\parahead{Breakdown.}
\label{section:End2EndBreakDown-66B}
We conduct the time breakdown of end-to-end inference with FT and \SYS{} for OPT-66B as shown in Fig.\ref{fig:Breakdown-Of-OPT-66B}.
Compared to FT-4GPU (FasterTransformer with 4 GPU used), \SYS{} with two GPUs can achieve lower normalized inference latency than FT-4GPU mainly because of 1) the reduction of MatMul time and 2) the reduction of cross-GPU communication overhead.
% Firstly, the MatMul time is reduced with \SYS{} kernels compared to using cuBLAS.
%Specifically, \SYS{} reduced the MatMul time from 44.6/45.0/45.1/46.5 seconds to 20.2/20.7/22.3/26.9 seconds at batch size 8/16/32/64.
%Besides, it reduces the cross-GPU communications from 8.6/11.0/13.6/17.8 seconds to 5.0/6.1/6.4/8.0 seconds at batch sizes 8/16/32/64.
%Compared to FT-2GPU, FT spends 42.1/42.5 seconds for MatMul computations at batch size 8/16, while our SpMM kernels only cost around 20.2/20.7 seconds.
%Put them all together, \SYS{} costs 38.3/42.7/48.6/62.1 seconds compared to FT-4GPU which costs 32.2/34.4/38.6/45.4 seconds respectively.

%\begin{figure}
%    \centering
%    \includegraphics[scale=0.3]{./Figures/AccumulatedKernelTime-OPT66B.pdf}
%    \caption{Breakdown of OPT-66B.}
%    \label{fig:Breakdown-Of-OPT-66B}
%\end{figure}

\subsubsection{Case Study: OPT-175B Model}

\begin{figure}
\centering
    \subfloat[Token Generation Throughput. \label{fig:OPT-175B-InferenceThroughput}]{
        \includegraphics[width=0.45\columnwidth]{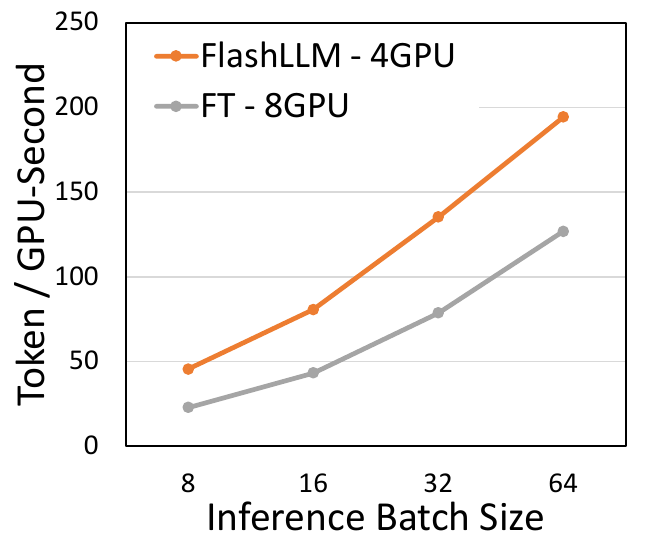}
    }
    \subfloat[Breakdown. \label{fig:Breakdown-Of-OPT-175B}]{
        \includegraphics[width=0.48\columnwidth]{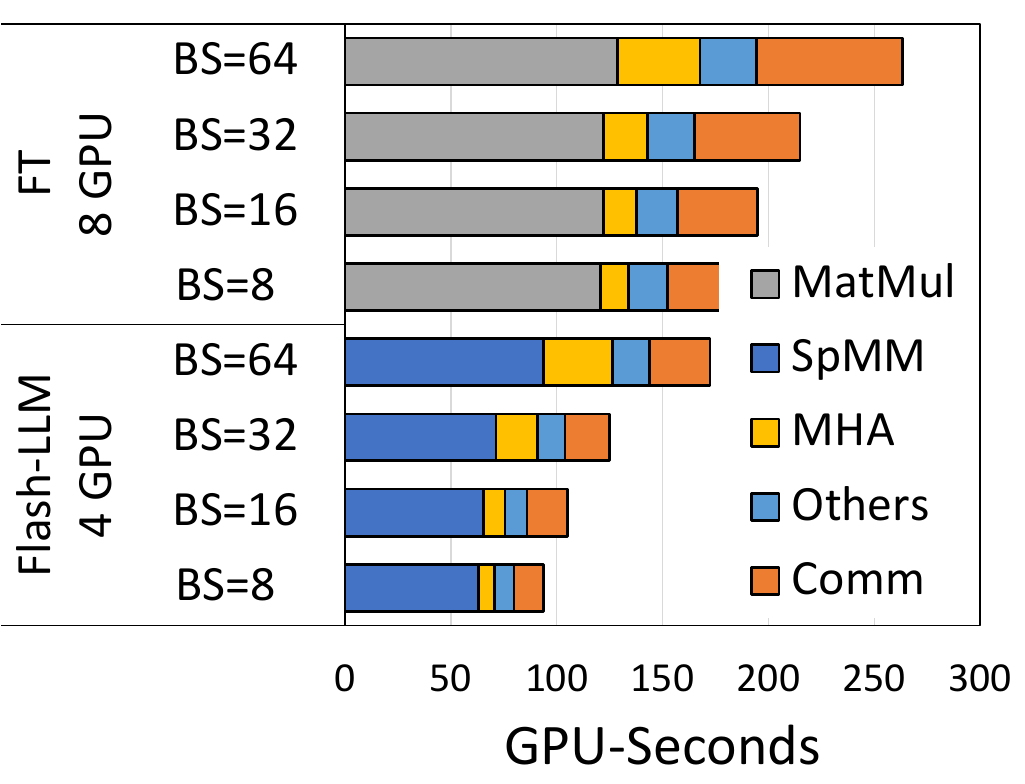}
    }
    % \vspace{-3ex}
    \caption{OPT-175B Inference}
    \label{fig:OPT-175B-Inference}
\end{figure}

\parahead{Results \& Breakdown}
We successfully run the inference of OPT-175B models with \SYS{} using 4 A100 GPUs.
In contrast, the weight of OPT-175B can not fit into 4 A100 GPUs with traditional solutions.
Thus, we do not show the performance of FT/DS using 4 GPUs here as they all run out of GPU memory.
In addition, we failed to run OPT-175B with DS using 8 GPUs.
Fig.\ref{fig:OPT-175B-InferenceThroughput} compares the performance of \SYS{} and FT where we only use 4 GPUs while FT uses 8 GPUs.
Compared to FT-8GPU, \SYS{} achieves $2.0\times$, $1.9\times$, $1.7\times$, and $1.5\times$ higher performance at batch sizes 8, 16, 32, and 64 respectively.
As shown in fig.\ref{fig:Breakdown-Of-OPT-175B}, both the MatMul time and the cross-GPU communication time are significantly reduced using \SYS{}.
%Compared to FT-8GPU, \SYS{} reduced the MatMul time from 120.7/122.1/122.1/128.6 seconds to 62.8/65.3/71.4/93.8 seconds at batch size 8/16/32/64.
%Besides, FT-8GPU cost 32.8/37.6/49.9/69.0 seconds for cross-GPU communications while \SYS{} with four GPUs only costs 13.7/19.0/21.0/28.4 seconds at batch sizes 8/16/32/64.

%\subsubsection{Cost Analysis}
%\TODO{Cost Analysis}
%Acure \& AWS
%In introduction or motivation section throw out the cost reduction, ROI

%Not only latency reduction + memory reduction, but also cost reduction

%quantization is more popular because they can be faster, sparse can not get speedup
\section{Related Work and Discussion}
\label{sec:related-work}

% \marginnote{
% % \footnotesize
% \cb{R3.M}
% }[-2.5in]

The parallel and distributed ML task execution is widely used for model training~\cite{ModelParallel, rajbhandari2020zero, Megatron-LM, jia2019beyond, GPipe, fan2021dapple, xianyan2022whale, xiaodong20optimizing, nie2023flexmoe, miao2021heterogeneity, zhang2022mics, miao2022galvatron, koliousis2019crossbow, li2022harmony, li13pytorch}.
With the growth of the model size, people start to support the LLM inference through tensor parallelism~\cite{Megatron-LM, jia2019beyond} to put parameters onto distributed devices~\cite{Faster-Transformer, DeepSpeed-Inference}.
However, the distributed execution of model inference introduces high communication costs and high economic investment.
Some works support LLM inference on a single GPU through memory offloading of data onto CPU memory and even disk~\cite{DeepSpeed-Inference, Flex-Gen}.
The offloading approach only works for latency non-sensitive applications (e.g. offline inference with large batch size), rather than the online inference tasks demanding very low latency.
In this work, we enable efficient LLM inference execution with fewer GPUs through efficient SpMM execution.

Model pruning is a common approach to reducing parameter numbers.
The structured sparsity is to enforce a structured distribution of non-zero elements during pruning, which usually could be friendly to hardware acceleration.
% Some work support to accelerate the structured SpMM on GPU Tensor Core.
NVIDIA Ampere GPU~\cite{Ampere_WhitePaper} supports 2:4 structured sparsity~\cite{SparseTensorCore-NVIDIA} to execute on tensor cores.
CuSPARSE\cite{cuSPARSE} can support structured SpMM on tensor core based on the Blocked-ELL~\cite{Blocked-ELL} format, which is a coarse-grained structured sparsity where model parameters are pruned in the granularity of a squared block (e.g. $32\times32$).
Some works prune the model parameters in the granularity of vectors to form the structured sparse format and make use of GPU tensor cores~\cite{TC-StructuredSparsity, TC-Quantized, Shfl-BW}.
While tensor cores can be enabled with certain structured sparsity, the major concern is that deep learning models pruned with structured sparsity usually suffer from more severe model accuracy degradation than unstructured sparsity
\cite{Wanda, Model_Compression, Deep_Compression, gomez2019learning, ullrich2017soft, SparseGPT}.

The unstructured sparsity is to prune the elements without forming a structured distribution, which is usually hard to accelerate on modern hardware architectures.
% \cb{
% \marginpar{\cb{R3.D2}}
STOREL \cite{OptimizingTensorPrograms} and TACO \cite{TACO} are CPU-based designs that can support SpMM with unstructured sparsity.
Instead, \SYS{} mainly focuses on the GPU-based SpMM with unstructured sparsity, as GPU is more widely used for large-scale deep-learning tasks since it usually has higher memory bandwidth, computational throughput, and energy efficiency.
% However, the mainstream of serving large deep-learning model inference in data centers is using GPUs instead, as GPUs are more suitable (higher memory bandwidth, computational throughput, and energy efficiency) for large-scale deep-learning tasks.
The typical approach to execute the unstructured SpMM on GPU is through SIMT cores, e.g. cuSPARSE\cite{cuSPARSE}, ASpT\cite{ASpT}, and Sputnik\cite{Sputnik}.
Under a moderate level of sparsity (<90\%), Sputnik significantly outperforms cuSPARSE and ASpT, but it struggles to beat its dense counterpart cuBLAS~\cite{cuBLAS} as it cannot utilize tensor cores.
% }
TC-GNN~\cite{TC-GNN} supports unstructured sparsity with tensor cores, which is customized for GNN where the sparsity ratio is extremely high (e.g. >99\%) and is not efficient for generative models requiring moderate-level sparsity.
SparTA~\cite{SparTA} proposes to utilize both sparse tensor cores~\cite{SparseTensorCore-NVIDIA} and SIMT cores to support unstructured sparsity by splitting the original sparse matrix into a 2:4 structured sparse matrix for tensor core execution (by cuSPARSELt~\cite{cuSPARSELt})) and an unstructured sparse matrix for SIMT core execution (by Sputnik\cite{Sputnik})).
However, if the sparse ratio is high, it has to excessively pad zeros to the 2:4 sparse matrix.
Besides, if there are too many non-zeros that can not meet the 2:4 requirement, the SIMT kernel would cause high latency and slow down the overall processing.
\SYS{} supports the moderate-level sparsity on tensor cores efficiently and does not require the 2:4-like distribution.
SparseTIR~\cite{SparseTIR} supports unstructured sparsity with tensor cores by splitting sparse matrices into $8 \times 1$ column vectors and omitting the vectors containing only zeros.
It can not work well under moderate sparsity (e.g., 80\%) as very few vectors can be skipped.
It can not outperform the dense baseline cuBLAS until the sparsity is higher than 95\%, while \SYS{} can outperform cuBLAS since 60\% sparsity.

The retraining-based pruning~\cite{liu2018rethinking, han2015learning} can achieve moderate-level sparsity with good accuracy, while the post-training pruning~\cite{SparseGPT} can only achieve quite low sparse ratios.
\SYS{} aims to optimize the SpMM with the moderate-level sparsity (e.g. 60\%-90\%) generated with the retraining-based pruning.
The retraining-based pruning usually consumes a high fine-tuning cost,
which as a result becomes one limitation of \SYS{}.

Model quantization~\cite{Model_Compression} is another approach to reduce the memory and computation for ML models, by transforming the data type into lower bits (e.g., 8-bits, 4-bits)~\cite{GPTQ, TC-Quantized}.
Model pruning and quantization are two orthogonal and complementary approaches to model compression.
This paper mainly focuses on supporting model pruning, which is orthogonal to model quantization.

\section{Conclusion}
\label{sec:conclusion}

We propose \SYS{}, a library for efficient large generative model inference through unstructured sparsity with tensor cores.
We observe that MatMuls in generative models inference are usually skinny, and are bounded by off-chip memory access.
We propose the \textit{Load-as-Sparse and Compute-as-Dense} approach for tensor core SpMM, 
reducing global memory footprint, addressing the memory access bottleneck, and tolerating redundant computations of skinny MatMuls.
We propose an effective software pipeline for unstructured SpMM with tensor cores, efficiently leveraging on-chip resources for sparse data extraction and coordinating sparse data extraction, dense data loading, and tensor core computation in an overlapped manner.
\SYS{} outperforms cuBLAS/Sputnik/SparTA by $1.4\times$/$3.6\times$/$1.4\times$, $1.7\times$/$3.0\times$/$1.4\times$, $2.1\times$/$2.0\times$/$1.6\times$ under 70\%, 80\% and 90\% sparsity.
We integrate \SYS{} kernels into Faster-Transformer for end-to-end generative model inference.
%The experiments on pruned OPT-30B, OPT-66B, and OPT-175B show that inference systems equipped with \SYS{} can achieve large generative model inference with higher inference throughput (from $1.41\times$ to $3.8\times$).
For \textit{tokens per GPU-second}, \SYS{} achieves up to $3.8 \times$ and $3.6 \times$ improvement over DeepSpeed and FasterTransformer on OPT-30B/66B/175B models with significantly lower inference cost.
%%% Main Chapters end %%%

%\begin{acks}
% This work was supported by the [...] Research Fund of [...] (Number [...]). Additional funding was provided by [...] and [...]. We also thank [...] for contributing [...].
%\end{acks}

%\clearpage
\balance
\bibliographystyle{ACM-Reference-Format}
\bibliography{sample}

%%% -*-BibTeX-*-
%%% Do NOT edit. File created by BibTeX with style
%%% ACM-Reference-Format-Journals [18-Jan-2012].

\begin{thebibliography}{65}

%%% ====================================================================
%%% NOTE TO THE USER: you can override these defaults by providing
%%% customized versions of any of these macros before the \bibliography
%%% command.  Each of them MUST provide its own final punctuation,
%%% except for \shownote{}, \showDOI{}, and \showURL{}.  The latter two
%%% do not use final punctuation, in order to avoid confusing it with
%%% the Web address.
%%%
%%% To suppress output of a particular field, define its macro to expand
%%% to an empty string, or better, \unskip, like this:
%%%
%%% \newcommand{\showDOI}[1]{\unskip}   % LaTeX syntax
%%%
%%% \def \showDOI #1{\unskip}           % plain TeX syntax
%%%
%%% ====================================================================

\ifx \showCODEN    \undefined \def \showCODEN     #1{\unskip}     \fi
\ifx \showDOI      \undefined \def \showDOI       #1{#1}\fi
\ifx \showISBNx    \undefined \def \showISBNx     #1{\unskip}     \fi
\ifx \showISBNxiii \undefined \def \showISBNxiii  #1{\unskip}     \fi
\ifx \showISSN     \undefined \def \showISSN      #1{\unskip}     \fi
\ifx \showLCCN     \undefined \def \showLCCN      #1{\unskip}     \fi
\ifx \shownote     \undefined \def \shownote      #1{#1}          \fi
\ifx \showarticletitle \undefined \def \showarticletitle #1{#1}   \fi
\ifx \showURL      \undefined \def \showURL       {\relax}        \fi
% The following commands are used for tagged output and should be
% invisible to TeX
\providecommand\bibfield[2]{#2}
\providecommand\bibinfo[2]{#2}
\providecommand\natexlab[1]{#1}
\providecommand\showeprint[2][]{arXiv:#2}

\bibitem[\protect\citeauthoryear{Aminabadi, Rajbhandari, Awan, Li, Li, Zheng, Ruwase, Smith, Zhang, Rasley, et~al\mbox{.}}{Aminabadi et~al\mbox{.}}{2022}]%
        {DeepSpeed-Inference}
\bibfield{author}{\bibinfo{person}{Reza~Yazdani Aminabadi}, \bibinfo{person}{Samyam Rajbhandari}, \bibinfo{person}{Ammar~Ahmad Awan}, \bibinfo{person}{Cheng Li}, \bibinfo{person}{Du Li}, \bibinfo{person}{Elton Zheng}, \bibinfo{person}{Olatunji Ruwase}, \bibinfo{person}{Shaden Smith}, \bibinfo{person}{Minjia Zhang}, \bibinfo{person}{Jeff Rasley}, {et~al\mbox{.}}} \bibinfo{year}{2022}\natexlab{}.
\newblock \showarticletitle{DeepSpeed-inference: enabling efficient inference of transformer models at unprecedented scale}. In \bibinfo{booktitle}{\emph{Proceedings of the International Conference on High Performance Computing, Networking, Storage and Analysis}}. \bibinfo{pages}{1--15}.
\newblock


\bibitem[\protect\citeauthoryear{Black, Biderman, Hallahan, Anthony, Gao, Golding, He, Leahy, McDonell, Phang, Pieler, Prashanth, Purohit, Reynolds, Tow, Wang, and Weinbach}{Black et~al\mbox{.}}{2022}]%
        {GPT-NEOX}
\bibfield{author}{\bibinfo{person}{Sid Black}, \bibinfo{person}{Stella Biderman}, \bibinfo{person}{Eric Hallahan}, \bibinfo{person}{Quentin Anthony}, \bibinfo{person}{Leo Gao}, \bibinfo{person}{Laurence Golding}, \bibinfo{person}{Horace He}, \bibinfo{person}{Connor Leahy}, \bibinfo{person}{Kyle McDonell}, \bibinfo{person}{Jason Phang}, \bibinfo{person}{Michael Pieler}, \bibinfo{person}{USVSN~Sai Prashanth}, \bibinfo{person}{Shivanshu Purohit}, \bibinfo{person}{Laria Reynolds}, \bibinfo{person}{Jonathan Tow}, \bibinfo{person}{Ben Wang}, {and} \bibinfo{person}{Samuel Weinbach}.} \bibinfo{year}{2022}\natexlab{}.
\newblock \showarticletitle{{GPT-NeoX-20B}: An Open-Source Autoregressive Language Model}. In \bibinfo{booktitle}{\emph{Proceedings of the ACL Workshop on Challenges \& Perspectives in Creating Large Language Models}}.
\newblock
\urldef\tempurl%
\url{https://arxiv.org/abs/2204.06745}
\showURL{%
\tempurl}


\bibitem[\protect\citeauthoryear{Brown, Mann, Ryder, Subbiah, Kaplan, Dhariwal, Neelakantan, Shyam, Sastry, Askell, et~al\mbox{.}}{Brown et~al\mbox{.}}{2020}]%
        {GPT-3}
\bibfield{author}{\bibinfo{person}{Tom Brown}, \bibinfo{person}{Benjamin Mann}, \bibinfo{person}{Nick Ryder}, \bibinfo{person}{Melanie Subbiah}, \bibinfo{person}{Jared~D Kaplan}, \bibinfo{person}{Prafulla Dhariwal}, \bibinfo{person}{Arvind Neelakantan}, \bibinfo{person}{Pranav Shyam}, \bibinfo{person}{Girish Sastry}, \bibinfo{person}{Amanda Askell}, {et~al\mbox{.}}} \bibinfo{year}{2020}\natexlab{}.
\newblock \showarticletitle{Language models are few-shot learners}.
\newblock \bibinfo{journal}{\emph{Advances in neural information processing systems}}  \bibinfo{volume}{33} (\bibinfo{year}{2020}), \bibinfo{pages}{1877--1901}.
\newblock


\bibitem[\protect\citeauthoryear{Chen, Qu, Liu, Ding, and Xie}{Chen et~al\mbox{.}}{2021}]%
        {TC-StructuredSparsity}
\bibfield{author}{\bibinfo{person}{Zhaodong Chen}, \bibinfo{person}{Zheng Qu}, \bibinfo{person}{Liu Liu}, \bibinfo{person}{Yufei Ding}, {and} \bibinfo{person}{Yuan Xie}.} \bibinfo{year}{2021}\natexlab{}.
\newblock \showarticletitle{Efficient tensor core-based GPU kernels for structured sparsity under reduced precision}. In \bibinfo{booktitle}{\emph{Proceedings of the International Conference for High Performance Computing, Networking, Storage and Analysis}}. \bibinfo{pages}{1--14}.
\newblock


\bibitem[\protect\citeauthoryear{EleutherAI}{EleutherAI}{2022}]%
        {GPT-NEOX-HF}
\bibfield{author}{\bibinfo{person}{EleutherAI}.} \bibinfo{year}{2022}\natexlab{}.
\newblock \bibinfo{booktitle}{\emph{GPT-NeoX-20B}}.
\newblock
\urldef\tempurl%
\url{https://huggingface.co/EleutherAI/gpt-neox-20b}
\showURL{%
\tempurl}


\bibitem[\protect\citeauthoryear{Face}{Face}{2023}]%
        {ModelParallel}
\bibfield{author}{\bibinfo{person}{Hugging Face}.} \bibinfo{year}{2023}\natexlab{}.
\newblock \bibinfo{booktitle}{\emph{Model Parallelism}}.
\newblock
\urldef\tempurl%
\url{https://huggingface.co/docs/transformers/v4.15.0/parallelism}
\showURL{%
\tempurl}


\bibitem[\protect\citeauthoryear{Fan, Rong, Meng, Cao, Wang, Zheng, Wu, Long, Yang, Xia, Diao, Liu, and Lin}{Fan et~al\mbox{.}}{2021}]%
        {fan2021dapple}
\bibfield{author}{\bibinfo{person}{Shiqing Fan}, \bibinfo{person}{Yi Rong}, \bibinfo{person}{Chen Meng}, \bibinfo{person}{Zongyan Cao}, \bibinfo{person}{Siyu Wang}, \bibinfo{person}{Zhen Zheng}, \bibinfo{person}{Chuan Wu}, \bibinfo{person}{Guoping Long}, \bibinfo{person}{Jun Yang}, \bibinfo{person}{Lixue Xia}, \bibinfo{person}{Lansong Diao}, \bibinfo{person}{Xiaoyong Liu}, {and} \bibinfo{person}{Wei Lin}.} \bibinfo{year}{2021}\natexlab{}.
\newblock \showarticletitle{DAPPLE: A pipelined data parallel approach for training large models}. In \bibinfo{booktitle}{\emph{Proceedings of the 26th ACM SIGPLAN Symposium on Principles and Practice of Parallel Programming}}. \bibinfo{pages}{431--445}.
\newblock


\bibitem[\protect\citeauthoryear{Frantar and Alistarh}{Frantar and Alistarh}{2023}]%
        {SparseGPT}
\bibfield{author}{\bibinfo{person}{Elias Frantar} {and} \bibinfo{person}{Dan Alistarh}.} \bibinfo{year}{2023}\natexlab{}.
\newblock \showarticletitle{Massive Language Models Can Be Accurately Pruned in One-Shot}.
\newblock \bibinfo{journal}{\emph{arXiv preprint arXiv:2301.00774}} (\bibinfo{year}{2023}).
\newblock


\bibitem[\protect\citeauthoryear{Frantar, Ashkboos, Hoefler, and Alistarh}{Frantar et~al\mbox{.}}{2022}]%
        {GPTQ}
\bibfield{author}{\bibinfo{person}{Elias Frantar}, \bibinfo{person}{Saleh Ashkboos}, \bibinfo{person}{Torsten Hoefler}, {and} \bibinfo{person}{Dan Alistarh}.} \bibinfo{year}{2022}\natexlab{}.
\newblock \showarticletitle{GPTQ: Accurate Post-Training Quantization for Generative Pre-trained Transformers}.
\newblock \bibinfo{journal}{\emph{arXiv preprint arXiv:2210.17323}} (\bibinfo{year}{2022}).
\newblock


\bibitem[\protect\citeauthoryear{Gale, Zaharia, Young, and Elsen}{Gale et~al\mbox{.}}{2020a}]%
        {Sputnik}
\bibfield{author}{\bibinfo{person}{Trevor Gale}, \bibinfo{person}{Matei Zaharia}, \bibinfo{person}{Cliff Young}, {and} \bibinfo{person}{Erich Elsen}.} \bibinfo{year}{2020}\natexlab{a}.
\newblock \showarticletitle{Sparse gpu kernels for deep learning}. In \bibinfo{booktitle}{\emph{SC20: International Conference for High Performance Computing, Networking, Storage and Analysis}}. IEEE, \bibinfo{pages}{1--14}.
\newblock


\bibitem[\protect\citeauthoryear{Gale, Zaharia, Young, and Elsen}{Gale et~al\mbox{.}}{2020b}]%
        {Sputnik-Github}
\bibfield{author}{\bibinfo{person}{Trevor Gale}, \bibinfo{person}{Matei Zaharia}, \bibinfo{person}{Cliff Young}, {and} \bibinfo{person}{Erich Elsen}.} \bibinfo{year}{2020}\natexlab{b}.
\newblock \bibinfo{booktitle}{\emph{sputnik github}}.
\newblock
\urldef\tempurl%
\url{https://github.com/google-research/sputnik/}
\showURL{%
\tempurl}


\bibitem[\protect\citeauthoryear{Gomez, Zhang, Kamalakara, Madaan, Swersky, Gal, and Hinton}{Gomez et~al\mbox{.}}{2019}]%
        {gomez2019learning}
\bibfield{author}{\bibinfo{person}{Aidan~N Gomez}, \bibinfo{person}{Ivan Zhang}, \bibinfo{person}{Siddhartha~Rao Kamalakara}, \bibinfo{person}{Divyam Madaan}, \bibinfo{person}{Kevin Swersky}, \bibinfo{person}{Yarin Gal}, {and} \bibinfo{person}{Geoffrey~E Hinton}.} \bibinfo{year}{2019}\natexlab{}.
\newblock \showarticletitle{Learning sparse networks using targeted dropout}.
\newblock \bibinfo{journal}{\emph{arXiv preprint arXiv:1905.13678}} (\bibinfo{year}{2019}).
\newblock


\bibitem[\protect\citeauthoryear{Gray, Radford, and Kingma}{Gray et~al\mbox{.}}{2017}]%
        {Blocked-ELL}
\bibfield{author}{\bibinfo{person}{Scott Gray}, \bibinfo{person}{Alec Radford}, {and} \bibinfo{person}{Diederik~P Kingma}.} \bibinfo{year}{2017}\natexlab{}.
\newblock \showarticletitle{Gpu kernels for block-sparse weights}.
\newblock \bibinfo{journal}{\emph{arXiv preprint arXiv:1711.09224}}  \bibinfo{volume}{3} (\bibinfo{year}{2017}), \bibinfo{pages}{2}.
\newblock


\bibitem[\protect\citeauthoryear{Han, Mao, and Dally}{Han et~al\mbox{.}}{2015a}]%
        {Deep_Compression}
\bibfield{author}{\bibinfo{person}{Song Han}, \bibinfo{person}{Huizi Mao}, {and} \bibinfo{person}{William~J Dally}.} \bibinfo{year}{2015}\natexlab{a}.
\newblock \showarticletitle{Deep compression: Compressing deep neural networks with pruning, trained quantization and huffman coding}.
\newblock \bibinfo{journal}{\emph{arXiv preprint arXiv:1510.00149}} (\bibinfo{year}{2015}).
\newblock


\bibitem[\protect\citeauthoryear{Han, Pool, Tran, and Dally}{Han et~al\mbox{.}}{2015b}]%
        {han2015learning}
\bibfield{author}{\bibinfo{person}{Song Han}, \bibinfo{person}{Jeff Pool}, \bibinfo{person}{John Tran}, {and} \bibinfo{person}{William Dally}.} \bibinfo{year}{2015}\natexlab{b}.
\newblock \showarticletitle{Learning both weights and connections for efficient neural network}.
\newblock \bibinfo{journal}{\emph{Advances in neural information processing systems}}  \bibinfo{volume}{28} (\bibinfo{year}{2015}).
\newblock


\bibitem[\protect\citeauthoryear{Hoefler, Alistarh, Ben-Nun, Dryden, and Peste}{Hoefler et~al\mbox{.}}{2021}]%
        {Model_Compression}
\bibfield{author}{\bibinfo{person}{Torsten Hoefler}, \bibinfo{person}{Dan Alistarh}, \bibinfo{person}{Tal Ben-Nun}, \bibinfo{person}{Nikoli Dryden}, {and} \bibinfo{person}{Alexandra Peste}.} \bibinfo{year}{2021}\natexlab{}.
\newblock \showarticletitle{Sparsity in Deep Learning: Pruning and growth for efficient inference and training in neural networks.}
\newblock \bibinfo{journal}{\emph{J. Mach. Learn. Res.}} \bibinfo{volume}{22}, \bibinfo{number}{241} (\bibinfo{year}{2021}), \bibinfo{pages}{1--124}.
\newblock


\bibitem[\protect\citeauthoryear{Hong, Sukumaran-Rajam, Nisa, Singh, and Sadayappan}{Hong et~al\mbox{.}}{2019}]%
        {ASpT}
\bibfield{author}{\bibinfo{person}{Changwan Hong}, \bibinfo{person}{Aravind Sukumaran-Rajam}, \bibinfo{person}{Israt Nisa}, \bibinfo{person}{Kunal Singh}, {and} \bibinfo{person}{P Sadayappan}.} \bibinfo{year}{2019}\natexlab{}.
\newblock \showarticletitle{Adaptive sparse tiling for sparse matrix multiplication}. In \bibinfo{booktitle}{\emph{Proceedings of the 24th Symposium on Principles and Practice of Parallel Programming}}. \bibinfo{pages}{300--314}.
\newblock


\bibitem[\protect\citeauthoryear{Huang, Li, Qin, Sun, Ding, and Xie}{Huang et~al\mbox{.}}{2022}]%
        {Shfl-BW}
\bibfield{author}{\bibinfo{person}{Guyue Huang}, \bibinfo{person}{Haoran Li}, \bibinfo{person}{Minghai Qin}, \bibinfo{person}{Fei Sun}, \bibinfo{person}{Yufei Ding}, {and} \bibinfo{person}{Yuan Xie}.} \bibinfo{year}{2022}\natexlab{}.
\newblock \showarticletitle{Shfl-BW: Accelerating Deep Neural Network Inference with Tensor-Core Aware Weight Pruning}. In \bibinfo{booktitle}{\emph{Proceedings of the 59th ACM/IEEE Design Automation Conference}} (San Francisco, California) \emph{(\bibinfo{series}{DAC '22})}. \bibinfo{publisher}{Association for Computing Machinery}, \bibinfo{address}{New York, NY, USA}, \bibinfo{pages}{1153–1158}.
\newblock
\showISBNx{9781450391429}
\urldef\tempurl%
\url{https://doi.org/10.1145/3489517.3530588}
\showDOI{\tempurl}


\bibitem[\protect\citeauthoryear{Huang, Cheng, Bapna, Firat, Chen, Chen, Lee, Ngiam, Le, Wu, et~al\mbox{.}}{Huang et~al\mbox{.}}{2019}]%
        {GPipe}
\bibfield{author}{\bibinfo{person}{Yanping Huang}, \bibinfo{person}{Youlong Cheng}, \bibinfo{person}{Ankur Bapna}, \bibinfo{person}{Orhan Firat}, \bibinfo{person}{Dehao Chen}, \bibinfo{person}{Mia Chen}, \bibinfo{person}{HyoukJoong Lee}, \bibinfo{person}{Jiquan Ngiam}, \bibinfo{person}{Quoc~V Le}, \bibinfo{person}{Yonghui Wu}, {et~al\mbox{.}}} \bibinfo{year}{2019}\natexlab{}.
\newblock \showarticletitle{Gpipe: Efficient training of giant neural networks using pipeline parallelism}.
\newblock \bibinfo{journal}{\emph{Advances in neural information processing systems}}  \bibinfo{volume}{32} (\bibinfo{year}{2019}).
\newblock


\bibitem[\protect\citeauthoryear{Jia, Jiang, Wang, Xiao, Shi, Zhang, Li, Chen, Li, Zheng, Liu, and Lin}{Jia et~al\mbox{.}}{2022}]%
        {xianyan2022whale}
\bibfield{author}{\bibinfo{person}{Xianyan Jia}, \bibinfo{person}{Le Jiang}, \bibinfo{person}{Ang Wang}, \bibinfo{person}{Wencong Xiao}, \bibinfo{person}{Ziji Shi}, \bibinfo{person}{Jie Zhang}, \bibinfo{person}{Xinyuan Li}, \bibinfo{person}{Langshi Chen}, \bibinfo{person}{Yong Li}, \bibinfo{person}{Zhen Zheng}, \bibinfo{person}{Xiaoyong Liu}, {and} \bibinfo{person}{Wei Lin}.} \bibinfo{year}{2022}\natexlab{}.
\newblock \showarticletitle{Whale: Efficient Giant Model Training over Heterogeneous GPUs}. In \bibinfo{booktitle}{\emph{2022 {USENIX} Annual Technical Conference, {USENIX} {ATC} 2022, Carlsbad, CA, USA, July 11-13, 2022}}, \bibfield{editor}{\bibinfo{person}{Jiri Schindler} {and} \bibinfo{person}{Noa Zilberman}} (Eds.). \bibinfo{publisher}{{USENIX} Association}, \bibinfo{pages}{673--688}.
\newblock


\bibitem[\protect\citeauthoryear{Jia, Zaharia, and Aiken}{Jia et~al\mbox{.}}{2019}]%
        {jia2019beyond}
\bibfield{author}{\bibinfo{person}{Zhihao Jia}, \bibinfo{person}{Matei Zaharia}, {and} \bibinfo{person}{Alex Aiken}.} \bibinfo{year}{2019}\natexlab{}.
\newblock \showarticletitle{Beyond Data and Model Parallelism for Deep Neural Networks.}
\newblock \bibinfo{journal}{\emph{Proceedings of Machine Learning and Systems}}  \bibinfo{volume}{1} (\bibinfo{year}{2019}), \bibinfo{pages}{1--13}.
\newblock


\bibitem[\protect\citeauthoryear{Kenton and Toutanova}{Kenton and Toutanova}{2019}]%
        {BERT}
\bibfield{author}{\bibinfo{person}{Jacob Devlin Ming-Wei~Chang Kenton} {and} \bibinfo{person}{Lee~Kristina Toutanova}.} \bibinfo{year}{2019}\natexlab{}.
\newblock \showarticletitle{BERT: Pre-training of Deep Bidirectional Transformers for Language Understanding}. In \bibinfo{booktitle}{\emph{Proceedings of NAACL-HLT}}. \bibinfo{pages}{4171--4186}.
\newblock


\bibitem[\protect\citeauthoryear{Kjolstad, Kamil, Chou, Lugato, and Amarasinghe}{Kjolstad et~al\mbox{.}}{2017}]%
        {TACO}
\bibfield{author}{\bibinfo{person}{Fredrik Kjolstad}, \bibinfo{person}{Shoaib Kamil}, \bibinfo{person}{Stephen Chou}, \bibinfo{person}{David Lugato}, {and} \bibinfo{person}{Saman Amarasinghe}.} \bibinfo{year}{2017}\natexlab{}.
\newblock \showarticletitle{The Tensor Algebra Compiler}.
\newblock \bibinfo{journal}{\emph{Proc. ACM Program. Lang.}} \bibinfo{volume}{1}, \bibinfo{number}{OOPSLA}, Article \bibinfo{articleno}{77} (\bibinfo{date}{oct} \bibinfo{year}{2017}), \bibinfo{numpages}{29}~pages.
\newblock
\urldef\tempurl%
\url{https://doi.org/10.1145/3133901}
\showDOI{\tempurl}


\bibitem[\protect\citeauthoryear{Koliousis, Watcharapichat, Weidlich, Mai, Costa, and Pietzuch}{Koliousis et~al\mbox{.}}{2019}]%
        {koliousis2019crossbow}
\bibfield{author}{\bibinfo{person}{Alexandros Koliousis}, \bibinfo{person}{Pijika Watcharapichat}, \bibinfo{person}{Matthias Weidlich}, \bibinfo{person}{Luo Mai}, \bibinfo{person}{Paolo Costa}, {and} \bibinfo{person}{Peter Pietzuch}.} \bibinfo{year}{2019}\natexlab{}.
\newblock \showarticletitle{CROSSBOW: Scaling Deep Learning with Small Batch Sizes on Multi-GPU Servers}.
\newblock \bibinfo{journal}{\emph{Proceedings of the VLDB Endowment}} \bibinfo{volume}{12}, \bibinfo{number}{11} (\bibinfo{year}{2019}).
\newblock


\bibitem[\protect\citeauthoryear{Li, Osawa, and Hoefler}{Li et~al\mbox{.}}{2022a}]%
        {TC-Quantized}
\bibfield{author}{\bibinfo{person}{Shigang Li}, \bibinfo{person}{Kazuki Osawa}, {and} \bibinfo{person}{Torsten Hoefler}.} \bibinfo{year}{2022}\natexlab{a}.
\newblock \showarticletitle{Efficient quantized sparse matrix operations on tensor cores}. In \bibinfo{booktitle}{\emph{Proceedings of the International Conference on High Performance Computing, Networking, Storage and Analysis}}. \bibinfo{pages}{1--15}.
\newblock


\bibitem[\protect\citeauthoryear{Li, Zhao, Varma, Salpekar, Noordhuis, Li, Paszke, Smith, Vaughan, Damania, et~al\mbox{.}}{Li et~al\mbox{.}}{2020}]%
        {li13pytorch}
\bibfield{author}{\bibinfo{person}{Shen Li}, \bibinfo{person}{Yanli Zhao}, \bibinfo{person}{Rohan Varma}, \bibinfo{person}{Omkar Salpekar}, \bibinfo{person}{Pieter Noordhuis}, \bibinfo{person}{Teng Li}, \bibinfo{person}{Adam Paszke}, \bibinfo{person}{Jeff Smith}, \bibinfo{person}{Brian Vaughan}, \bibinfo{person}{Pritam Damania}, {et~al\mbox{.}}} \bibinfo{year}{2020}\natexlab{}.
\newblock \showarticletitle{PyTorch Distributed: Experiences on Accelerating Data Parallel Training}.
\newblock \bibinfo{journal}{\emph{Proceedings of the VLDB Endowment}} \bibinfo{volume}{13}, \bibinfo{number}{12} (\bibinfo{year}{2020}).
\newblock


\bibitem[\protect\citeauthoryear{Li, Phanishayee, Murray, Tarnawski, and Kim}{Li et~al\mbox{.}}{2022b}]%
        {li2022harmony}
\bibfield{author}{\bibinfo{person}{Youjie Li}, \bibinfo{person}{Amar Phanishayee}, \bibinfo{person}{Derek Murray}, \bibinfo{person}{Jakub Tarnawski}, {and} \bibinfo{person}{Nam~Sung Kim}.} \bibinfo{year}{2022}\natexlab{b}.
\newblock \showarticletitle{Harmony: Overcoming the Hurdles of GPU Memory Capacity to Train Massive DNN Models on Commodity Servers}.
\newblock \bibinfo{journal}{\emph{Proc. VLDB Endow.}} \bibinfo{volume}{15}, \bibinfo{number}{11} (\bibinfo{date}{jul} \bibinfo{year}{2022}), \bibinfo{pages}{2747–2760}.
\newblock
\showISSN{2150-8097}


\bibitem[\protect\citeauthoryear{Lin, Zhang, Li, Chen, Chao, Wang, Li, Tian, and Ji}{Lin et~al\mbox{.}}{2022}]%
        {SparsityMethod}
\bibfield{author}{\bibinfo{person}{Mingbao Lin}, \bibinfo{person}{Yuxin Zhang}, \bibinfo{person}{Yuchao Li}, \bibinfo{person}{Bohong Chen}, \bibinfo{person}{Fei Chao}, \bibinfo{person}{Mengdi Wang}, \bibinfo{person}{Shen Li}, \bibinfo{person}{Yonghong Tian}, {and} \bibinfo{person}{Rongrong Ji}.} \bibinfo{year}{2022}\natexlab{}.
\newblock \showarticletitle{1xn pattern for pruning convolutional neural networks}.
\newblock \bibinfo{journal}{\emph{IEEE Transactions on Pattern Analysis and Machine Intelligence}} (\bibinfo{year}{2022}).
\newblock


\bibitem[\protect\citeauthoryear{Liu, Sun, Zhou, Huang, and Darrell}{Liu et~al\mbox{.}}{2018}]%
        {liu2018rethinking}
\bibfield{author}{\bibinfo{person}{Zhuang Liu}, \bibinfo{person}{Mingjie Sun}, \bibinfo{person}{Tinghui Zhou}, \bibinfo{person}{Gao Huang}, {and} \bibinfo{person}{Trevor Darrell}.} \bibinfo{year}{2018}\natexlab{}.
\newblock \showarticletitle{Rethinking the value of network pruning}.
\newblock \bibinfo{journal}{\emph{arXiv preprint arXiv:1810.05270}} (\bibinfo{year}{2018}).
\newblock


\bibitem[\protect\citeauthoryear{Miao, Nie, Shao, Yang, Jiang, Ma, and Cui}{Miao et~al\mbox{.}}{2021}]%
        {miao2021heterogeneity}
\bibfield{author}{\bibinfo{person}{Xupeng Miao}, \bibinfo{person}{Xiaonan Nie}, \bibinfo{person}{Yingxia Shao}, \bibinfo{person}{Zhi Yang}, \bibinfo{person}{Jiawei Jiang}, \bibinfo{person}{Lingxiao Ma}, {and} \bibinfo{person}{Bin Cui}.} \bibinfo{year}{2021}\natexlab{}.
\newblock \showarticletitle{Heterogeneity-aware distributed machine learning training via partial reduce}. In \bibinfo{booktitle}{\emph{Proceedings of the 2021 International Conference on Management of Data}}. \bibinfo{pages}{2262--2270}.
\newblock


\bibitem[\protect\citeauthoryear{Miao, Wang, Jiang, Shi, Nie, Zhang, and Cui}{Miao et~al\mbox{.}}{2022}]%
        {miao2022galvatron}
\bibfield{author}{\bibinfo{person}{Xupeng Miao}, \bibinfo{person}{Yujie Wang}, \bibinfo{person}{Youhe Jiang}, \bibinfo{person}{Chunan Shi}, \bibinfo{person}{Xiaonan Nie}, \bibinfo{person}{Hailin Zhang}, {and} \bibinfo{person}{Bin Cui}.} \bibinfo{year}{2022}\natexlab{}.
\newblock \showarticletitle{Galvatron: Efficient Transformer Training over Multiple GPUs Using Automatic Parallelism}.
\newblock \bibinfo{journal}{\emph{Proc. VLDB Endow.}} \bibinfo{volume}{16}, \bibinfo{number}{3} (\bibinfo{date}{nov} \bibinfo{year}{2022}), \bibinfo{pages}{470–479}.
\newblock
\showISSN{2150-8097}


\bibitem[\protect\citeauthoryear{Mishra, Latorre, Pool, Stosic, Stosic, Venkatesh, Yu, and Micikevicius}{Mishra et~al\mbox{.}}{2021}]%
        {SparseTensorCore-NVIDIA}
\bibfield{author}{\bibinfo{person}{Asit Mishra}, \bibinfo{person}{Jorge~Albericio Latorre}, \bibinfo{person}{Jeff Pool}, \bibinfo{person}{Darko Stosic}, \bibinfo{person}{Dusan Stosic}, \bibinfo{person}{Ganesh Venkatesh}, \bibinfo{person}{Chong Yu}, {and} \bibinfo{person}{Paulius Micikevicius}.} \bibinfo{year}{2021}\natexlab{}.
\newblock \bibinfo{title}{Accelerating Sparse Deep Neural Networks}.
\newblock
\newblock
\showeprint[arxiv]{2104.08378}~[cs.LG]


\bibitem[\protect\citeauthoryear{Molchanov, Mallya, Tyree, Frosio, and Kautz}{Molchanov et~al\mbox{.}}{2019}]%
        {TaylorPruning}
\bibfield{author}{\bibinfo{person}{Pavlo Molchanov}, \bibinfo{person}{Arun Mallya}, \bibinfo{person}{Stephen Tyree}, \bibinfo{person}{Iuri Frosio}, {and} \bibinfo{person}{Jan Kautz}.} \bibinfo{year}{2019}\natexlab{}.
\newblock \showarticletitle{Importance estimation for neural network pruning}. In \bibinfo{booktitle}{\emph{Proceedings of the IEEE/CVF conference on computer vision and pattern recognition}}. \bibinfo{pages}{11264--11272}.
\newblock


\bibitem[\protect\citeauthoryear{Narayan, Chami, Orr, and R\'{e}}{Narayan et~al\mbox{.}}{2022}]%
        {10.14778/3574245.3574258}
\bibfield{author}{\bibinfo{person}{Avanika Narayan}, \bibinfo{person}{Ines Chami}, \bibinfo{person}{Laurel Orr}, {and} \bibinfo{person}{Christopher R\'{e}}.} \bibinfo{year}{2022}\natexlab{}.
\newblock \showarticletitle{Can Foundation Models Wrangle Your Data?}
\newblock \bibinfo{journal}{\emph{Proc. VLDB Endow.}} \bibinfo{volume}{16}, \bibinfo{number}{4} (\bibinfo{date}{dec} \bibinfo{year}{2022}), \bibinfo{pages}{738–746}.
\newblock
\showISSN{2150-8097}


\bibitem[\protect\citeauthoryear{Nie, Miao, Wang, Yang, Xue, Ma, Cao, and Cui}{Nie et~al\mbox{.}}{2023}]%
        {nie2023flexmoe}
\bibfield{author}{\bibinfo{person}{Xiaonan Nie}, \bibinfo{person}{Xupeng Miao}, \bibinfo{person}{Zilong Wang}, \bibinfo{person}{Zichao Yang}, \bibinfo{person}{Jilong Xue}, \bibinfo{person}{Lingxiao Ma}, \bibinfo{person}{Gang Cao}, {and} \bibinfo{person}{Bin Cui}.} \bibinfo{year}{2023}\natexlab{}.
\newblock \showarticletitle{FlexMoE: Scaling Large-scale Sparse Pre-trained Model Training via Dynamic Device Placement}.
\newblock \bibinfo{journal}{\emph{arXiv preprint arXiv:2304.03946}} (\bibinfo{year}{2023}).
\newblock


\bibitem[\protect\citeauthoryear{NVIDIA}{NVIDIA}{2020}]%
        {Ampere_WhitePaper}
\bibfield{author}{\bibinfo{person}{NVIDIA}.} \bibinfo{year}{2020}\natexlab{}.
\newblock \bibinfo{booktitle}{\emph{NVIDIA A100 Tensor Core GPU Architecture}}.
\newblock
\urldef\tempurl%
\url{https://images.nvidia.com/aem-dam/en-zz/Solutions/data-center/nvidia-ampere-architecture-whitepaper.pdf}
\showURL{%
\tempurl}


\bibitem[\protect\citeauthoryear{NVIDIA}{NVIDIA}{2022a}]%
        {Faster-Transformer}
\bibfield{author}{\bibinfo{person}{NVIDIA}.} \bibinfo{year}{2022}\natexlab{a}.
\newblock \bibinfo{booktitle}{\emph{NVIDIA Faster-Transformer}}.
\newblock
\urldef\tempurl%
\url{https://github.com/NVIDIA/FasterTransformer}
\showURL{%
\tempurl}


\bibitem[\protect\citeauthoryear{NVIDIA}{NVIDIA}{2022b}]%
        {Hopper_WhitePaper}
\bibfield{author}{\bibinfo{person}{NVIDIA}.} \bibinfo{year}{2022}\natexlab{b}.
\newblock \bibinfo{booktitle}{\emph{NVIDIA H100 Tensor Core GPU Architecture}}.
\newblock
\urldef\tempurl%
\url{https://www.hpctech.co.jp/catalog/gtc22-whitepaper-hopper_v1.01.pdf}
\showURL{%
\tempurl}


\bibitem[\protect\citeauthoryear{NVIDIA}{NVIDIA}{2023a}]%
        {cuBLAS}
\bibfield{author}{\bibinfo{person}{NVIDIA}.} \bibinfo{year}{2023}\natexlab{a}.
\newblock \bibinfo{booktitle}{\emph{cuBLAS Docs}}.
\newblock
\urldef\tempurl%
\url{https://docs.nvidia.com/cuda/cublas/index.html}
\showURL{%
\tempurl}


\bibitem[\protect\citeauthoryear{NVIDIA}{NVIDIA}{2023b}]%
        {cuSPARSE}
\bibfield{author}{\bibinfo{person}{NVIDIA}.} \bibinfo{year}{2023}\natexlab{b}.
\newblock \bibinfo{booktitle}{\emph{cuSPARSE Library}}.
\newblock
\urldef\tempurl%
\url{https://docs.nvidia.com/cuda/cusparse/index.html}
\showURL{%
\tempurl}


\bibitem[\protect\citeauthoryear{NVIDIA}{NVIDIA}{2023c}]%
        {cuSPARSELt}
\bibfield{author}{\bibinfo{person}{NVIDIA}.} \bibinfo{year}{2023}\natexlab{c}.
\newblock \bibinfo{booktitle}{\emph{cuSPARSELt Library}}.
\newblock
\urldef\tempurl%
\url{https://docs.nvidia.com/cuda/cusparselt/}
\showURL{%
\tempurl}


\bibitem[\protect\citeauthoryear{NVIDIA}{NVIDIA}{2023d}]%
        {cutlass}
\bibfield{author}{\bibinfo{person}{NVIDIA}.} \bibinfo{year}{2023}\natexlab{d}.
\newblock \bibinfo{booktitle}{\emph{CUTLASS 3.2}}.
\newblock
\urldef\tempurl%
\url{https://github.com/NVIDIA/cutlass}
\showURL{%
\tempurl}


\bibitem[\protect\citeauthoryear{NVIDIA}{NVIDIA}{2023e}]%
        {NsightCompute}
\bibfield{author}{\bibinfo{person}{NVIDIA}.} \bibinfo{year}{2023}\natexlab{e}.
\newblock \bibinfo{booktitle}{\emph{Nsight Compute Profiling Guide}}.
\newblock
\urldef\tempurl%
\url{https://docs.nvidia.com/nsight-compute/ProfilingGuide/#introduction}
\showURL{%
\tempurl}


\bibitem[\protect\citeauthoryear{NVIDIA}{NVIDIA}{2023f}]%
        {NsightSystem}
\bibfield{author}{\bibinfo{person}{NVIDIA}.} \bibinfo{year}{2023}\natexlab{f}.
\newblock \bibinfo{booktitle}{\emph{Nsight System}}.
\newblock
\urldef\tempurl%
\url{https://developer.nvidia.com/nsight-systems}
\showURL{%
\tempurl}


\bibitem[\protect\citeauthoryear{Radford, Wu, Child, Luan, Amodei, Sutskever, et~al\mbox{.}}{Radford et~al\mbox{.}}{2019}]%
        {GPT-2}
\bibfield{author}{\bibinfo{person}{Alec Radford}, \bibinfo{person}{Jeffrey Wu}, \bibinfo{person}{Rewon Child}, \bibinfo{person}{David Luan}, \bibinfo{person}{Dario Amodei}, \bibinfo{person}{Ilya Sutskever}, {et~al\mbox{.}}} \bibinfo{year}{2019}\natexlab{}.
\newblock \showarticletitle{Language models are unsupervised multitask learners}.
\newblock \bibinfo{journal}{\emph{OpenAI blog}} \bibinfo{volume}{1}, \bibinfo{number}{8} (\bibinfo{year}{2019}), \bibinfo{pages}{9}.
\newblock


\bibitem[\protect\citeauthoryear{Rajbhandari, Rasley, Ruwase, and He}{Rajbhandari et~al\mbox{.}}{2020}]%
        {rajbhandari2020zero}
\bibfield{author}{\bibinfo{person}{Samyam Rajbhandari}, \bibinfo{person}{Jeff Rasley}, \bibinfo{person}{Olatunji Ruwase}, {and} \bibinfo{person}{Yuxiong He}.} \bibinfo{year}{2020}\natexlab{}.
\newblock \showarticletitle{Zero: Memory optimizations toward training trillion parameter models}. In \bibinfo{booktitle}{\emph{SC20: International Conference for High Performance Computing, Networking, Storage and Analysis}}. IEEE, \bibinfo{pages}{1--16}.
\newblock


\bibitem[\protect\citeauthoryear{Schleich, Shaikhha, and Suciu}{Schleich et~al\mbox{.}}{2023}]%
        {OptimizingTensorPrograms}
\bibfield{author}{\bibinfo{person}{Maximilian Schleich}, \bibinfo{person}{Amir Shaikhha}, {and} \bibinfo{person}{Dan Suciu}.} \bibinfo{year}{2023}\natexlab{}.
\newblock \showarticletitle{Optimizing Tensor Programs on Flexible Storage}.
\newblock \bibinfo{journal}{\emph{Proceedings of the ACM on Management of Data}} \bibinfo{volume}{1}, \bibinfo{number}{1} (\bibinfo{year}{2023}), \bibinfo{pages}{1--27}.
\newblock


\bibitem[\protect\citeauthoryear{Sheng, Zheng, Yuan, Li, Ryabinin, Fu, Xie, Chen, Barrett, Gonzalez, et~al\mbox{.}}{Sheng et~al\mbox{.}}{2023}]%
        {Flex-Gen}
\bibfield{author}{\bibinfo{person}{Ying Sheng}, \bibinfo{person}{Lianmin Zheng}, \bibinfo{person}{Binhang Yuan}, \bibinfo{person}{Zhuohan Li}, \bibinfo{person}{Max Ryabinin}, \bibinfo{person}{Daniel~Y Fu}, \bibinfo{person}{Zhiqiang Xie}, \bibinfo{person}{Beidi Chen}, \bibinfo{person}{Clark Barrett}, \bibinfo{person}{Joseph~E Gonzalez}, {et~al\mbox{.}}} \bibinfo{year}{2023}\natexlab{}.
\newblock \showarticletitle{High-throughput generative inference of large language models with a single gpu}.
\newblock \bibinfo{journal}{\emph{arXiv preprint arXiv:2303.06865}} (\bibinfo{year}{2023}).
\newblock


\bibitem[\protect\citeauthoryear{Shoeybi, Patwary, Puri, LeGresley, Casper, and Catanzaro}{Shoeybi et~al\mbox{.}}{2019}]%
        {Megatron-LM}
\bibfield{author}{\bibinfo{person}{Mohammad Shoeybi}, \bibinfo{person}{Mostofa Patwary}, \bibinfo{person}{Raul Puri}, \bibinfo{person}{Patrick LeGresley}, \bibinfo{person}{Jared Casper}, {and} \bibinfo{person}{Bryan Catanzaro}.} \bibinfo{year}{2019}\natexlab{}.
\newblock \showarticletitle{Megatron-lm: Training multi-billion parameter language models using model parallelism}.
\newblock \bibinfo{journal}{\emph{arXiv preprint arXiv:1909.08053}} (\bibinfo{year}{2019}).
\newblock


\bibitem[\protect\citeauthoryear{Smith, Patwary, Norick, LeGresley, Rajbhandari, Casper, Liu, Prabhumoye, Zerveas, Korthikanti, Zhang, Child, Aminabadi, Bernauer, Song, Shoeybi, He, Houston, Tiwary, and Catanzaro}{Smith et~al\mbox{.}}{2022}]%
        {Megatron-Turing-NLG}
\bibfield{author}{\bibinfo{person}{Shaden Smith}, \bibinfo{person}{Mostofa Patwary}, \bibinfo{person}{Brandon Norick}, \bibinfo{person}{Patrick LeGresley}, \bibinfo{person}{Samyam Rajbhandari}, \bibinfo{person}{Jared Casper}, \bibinfo{person}{Zhun Liu}, \bibinfo{person}{Shrimai Prabhumoye}, \bibinfo{person}{George Zerveas}, \bibinfo{person}{Vijay Korthikanti}, \bibinfo{person}{Elton Zhang}, \bibinfo{person}{Rewon Child}, \bibinfo{person}{Reza~Yazdani Aminabadi}, \bibinfo{person}{Julie Bernauer}, \bibinfo{person}{Xia Song}, \bibinfo{person}{Mohammad Shoeybi}, \bibinfo{person}{Yuxiong He}, \bibinfo{person}{Michael Houston}, \bibinfo{person}{Saurabh Tiwary}, {and} \bibinfo{person}{Bryan Catanzaro}.} \bibinfo{year}{2022}\natexlab{}.
\newblock \bibinfo{title}{Using DeepSpeed and Megatron to Train Megatron-Turing NLG 530B, A Large-Scale Generative Language Model}.
\newblock
\newblock
\showeprint[arxiv]{2201.11990}~[cs.CL]


\bibitem[\protect\citeauthoryear{Sun, Liu, Bair, and Kolter}{Sun et~al\mbox{.}}{2023}]%
        {Wanda}
\bibfield{author}{\bibinfo{person}{Mingjie Sun}, \bibinfo{person}{Zhuang Liu}, \bibinfo{person}{Anna Bair}, {and} \bibinfo{person}{J.~Zico Kolter}.} \bibinfo{year}{2023}\natexlab{}.
\newblock \bibinfo{title}{A Simple and Effective Pruning Approach for Large Language Models}.
\newblock
\newblock
\showeprint[arxiv]{2306.11695}~[cs.CL]


\bibitem[\protect\citeauthoryear{Trummer}{Trummer}{2022}]%
        {10.14778/3554821.3554896}
\bibfield{author}{\bibinfo{person}{Immanuel Trummer}.} \bibinfo{year}{2022}\natexlab{}.
\newblock \showarticletitle{From BERT to GPT-3 Codex: Harnessing the Potential of Very Large Language Models for Data Management}.
\newblock \bibinfo{journal}{\emph{Proc. VLDB Endow.}} \bibinfo{volume}{15}, \bibinfo{number}{12} (\bibinfo{date}{aug} \bibinfo{year}{2022}), \bibinfo{pages}{3770–3773}.
\newblock
\showISSN{2150-8097}


\bibitem[\protect\citeauthoryear{Tuli, Casale, and Jennings}{Tuli et~al\mbox{.}}{2022}]%
        {10.14778/3514061.3514067}
\bibfield{author}{\bibinfo{person}{Shreshth Tuli}, \bibinfo{person}{Giuliano Casale}, {and} \bibinfo{person}{Nicholas~R. Jennings}.} \bibinfo{year}{2022}\natexlab{}.
\newblock \showarticletitle{TranAD: Deep Transformer Networks for Anomaly Detection in Multivariate Time Series Data}.
\newblock \bibinfo{journal}{\emph{Proc. VLDB Endow.}} \bibinfo{volume}{15}, \bibinfo{number}{6} (\bibinfo{date}{feb} \bibinfo{year}{2022}), \bibinfo{pages}{1201–1214}.
\newblock
\showISSN{2150-8097}


\bibitem[\protect\citeauthoryear{Ullrich, Meeds, and Welling}{Ullrich et~al\mbox{.}}{2017}]%
        {ullrich2017soft}
\bibfield{author}{\bibinfo{person}{Karen Ullrich}, \bibinfo{person}{Edward Meeds}, {and} \bibinfo{person}{Max Welling}.} \bibinfo{year}{2017}\natexlab{}.
\newblock \showarticletitle{Soft weight-sharing for neural network compression}.
\newblock \bibinfo{journal}{\emph{arXiv preprint arXiv:1702.04008}} (\bibinfo{year}{2017}).
\newblock


\bibitem[\protect\citeauthoryear{Vaswani, Shazeer, Parmar, Uszkoreit, Jones, Gomez, Kaiser, and Polosukhin}{Vaswani et~al\mbox{.}}{2017}]%
        {Attention_Is_All_You_Need}
\bibfield{author}{\bibinfo{person}{Ashish Vaswani}, \bibinfo{person}{Noam Shazeer}, \bibinfo{person}{Niki Parmar}, \bibinfo{person}{Jakob Uszkoreit}, \bibinfo{person}{Llion Jones}, \bibinfo{person}{Aidan~N Gomez}, \bibinfo{person}{{\L}ukasz Kaiser}, {and} \bibinfo{person}{Illia Polosukhin}.} \bibinfo{year}{2017}\natexlab{}.
\newblock \showarticletitle{Attention is all you need}.
\newblock \bibinfo{journal}{\emph{Advances in neural information processing systems}}  \bibinfo{volume}{30} (\bibinfo{year}{2017}).
\newblock


\bibitem[\protect\citeauthoryear{Wang, Pruksachatkun, Nangia, Singh, Michael, Hill, Levy, and Bowman}{Wang et~al\mbox{.}}{2019}]%
        {SuperGlue}
\bibfield{author}{\bibinfo{person}{Alex Wang}, \bibinfo{person}{Yada Pruksachatkun}, \bibinfo{person}{Nikita Nangia}, \bibinfo{person}{Amanpreet Singh}, \bibinfo{person}{Julian Michael}, \bibinfo{person}{Felix Hill}, \bibinfo{person}{Omer Levy}, {and} \bibinfo{person}{Samuel Bowman}.} \bibinfo{year}{2019}\natexlab{}.
\newblock \showarticletitle{Superglue: A stickier benchmark for general-purpose language understanding systems}.
\newblock \bibinfo{journal}{\emph{Advances in neural information processing systems}}  \bibinfo{volume}{32} (\bibinfo{year}{2019}).
\newblock


\bibitem[\protect\citeauthoryear{Wang, Feng, Wang, and Ding}{Wang et~al\mbox{.}}{2023}]%
        {TC-GNN}
\bibfield{author}{\bibinfo{person}{Yuke Wang}, \bibinfo{person}{Boyuan Feng}, \bibinfo{person}{Zheng Wang}, {and} \bibinfo{person}{Yufei Ding}.} \bibinfo{year}{2023}\natexlab{}.
\newblock \bibinfo{title}{TC-GNN: Accelerating Sparse Graph Neural Network Computation Via Dense Tensor Core on GPUs}.
\newblock
\newblock
\showeprint[arxiv]{2112.02052}~[cs.LG]


\bibitem[\protect\citeauthoryear{Williams, Waterman, and Patterson}{Williams et~al\mbox{.}}{2009}]%
        {Roofline_Model}
\bibfield{author}{\bibinfo{person}{Samuel Williams}, \bibinfo{person}{Andrew Waterman}, {and} \bibinfo{person}{David Patterson}.} \bibinfo{year}{2009}\natexlab{}.
\newblock \showarticletitle{Roofline: an insightful visual performance model for multicore architectures}.
\newblock \bibinfo{journal}{\emph{Commun. ACM}} \bibinfo{volume}{52}, \bibinfo{number}{4} (\bibinfo{year}{2009}), \bibinfo{pages}{65--76}.
\newblock


\bibitem[\protect\citeauthoryear{Ye, Lai, Shao, Chen, and Ceze}{Ye et~al\mbox{.}}{2023}]%
        {SparseTIR}
\bibfield{author}{\bibinfo{person}{Zihao Ye}, \bibinfo{person}{Ruihang Lai}, \bibinfo{person}{Junru Shao}, \bibinfo{person}{Tianqi Chen}, {and} \bibinfo{person}{Luis Ceze}.} \bibinfo{year}{2023}\natexlab{}.
\newblock \showarticletitle{SparseTIR: Composable abstractions for sparse compilation in deep learning}. In \bibinfo{booktitle}{\emph{Proceedings of the 28th ACM International Conference on Architectural Support for Programming Languages and Operating Systems, Volume 3}}. \bibinfo{pages}{660--678}.
\newblock


\bibitem[\protect\citeauthoryear{Yi, Zhang, Luo, Long, Diao, Wu, Zheng, Yang, and Lin}{Yi et~al\mbox{.}}{2020}]%
        {xiaodong20optimizing}
\bibfield{author}{\bibinfo{person}{Xiaodong Yi}, \bibinfo{person}{Shiwei Zhang}, \bibinfo{person}{Ziyue Luo}, \bibinfo{person}{Guoping Long}, \bibinfo{person}{Lansong Diao}, \bibinfo{person}{Chuan Wu}, \bibinfo{person}{Zhen Zheng}, \bibinfo{person}{Jun Yang}, {and} \bibinfo{person}{Wei Lin}.} \bibinfo{year}{2020}\natexlab{}.
\newblock \showarticletitle{Optimizing distributed training deployment in heterogeneous {GPU} clusters}. In \bibinfo{booktitle}{\emph{CoNEXT '20: The 16th International Conference on emerging Networking EXperiments and Technologies, Barcelona, Spain, December, 2020}}, \bibfield{editor}{\bibinfo{person}{Dongsu Han} {and} \bibinfo{person}{Anja Feldmann}} (Eds.). \bibinfo{publisher}{{ACM}}, \bibinfo{pages}{93--107}.
\newblock


\bibitem[\protect\citeauthoryear{Zhang, Roller, Goyal, Artetxe, Chen, Chen, Dewan, Diab, Li, Lin, Mihaylov, Ott, Shleifer, Shuster, Simig, Koura, Sridhar, Wang, and Zettlemoyer}{Zhang et~al\mbox{.}}{2022a}]%
        {OPT-Models}
\bibfield{author}{\bibinfo{person}{Susan Zhang}, \bibinfo{person}{Stephen Roller}, \bibinfo{person}{Naman Goyal}, \bibinfo{person}{Mikel Artetxe}, \bibinfo{person}{Moya Chen}, \bibinfo{person}{Shuohui Chen}, \bibinfo{person}{Christopher Dewan}, \bibinfo{person}{Mona Diab}, \bibinfo{person}{Xian Li}, \bibinfo{person}{Xi~Victoria Lin}, \bibinfo{person}{Todor Mihaylov}, \bibinfo{person}{Myle Ott}, \bibinfo{person}{Sam Shleifer}, \bibinfo{person}{Kurt Shuster}, \bibinfo{person}{Daniel Simig}, \bibinfo{person}{Punit~Singh Koura}, \bibinfo{person}{Anjali Sridhar}, \bibinfo{person}{Tianlu Wang}, {and} \bibinfo{person}{Luke Zettlemoyer}.} \bibinfo{year}{2022}\natexlab{a}.
\newblock \bibinfo{title}{OPT: Open Pre-trained Transformer Language Models}.
\newblock
\newblock
\showeprint[arxiv]{2205.01068}~[cs.CL]


\bibitem[\protect\citeauthoryear{Zhang, Zheng, Wang, Chiu, Karypis, Chilimbi, Li, and Jin}{Zhang et~al\mbox{.}}{2022b}]%
        {zhang2022mics}
\bibfield{author}{\bibinfo{person}{Zhen Zhang}, \bibinfo{person}{Shuai Zheng}, \bibinfo{person}{Yida Wang}, \bibinfo{person}{Justin Chiu}, \bibinfo{person}{George Karypis}, \bibinfo{person}{Trishul Chilimbi}, \bibinfo{person}{Mu Li}, {and} \bibinfo{person}{Xin Jin}.} \bibinfo{year}{2022}\natexlab{b}.
\newblock \showarticletitle{MiCS: near-linear scaling for training gigantic model on public cloud}.
\newblock \bibinfo{journal}{\emph{Proceedings of the VLDB Endowment}} \bibinfo{volume}{16}, \bibinfo{number}{1} (\bibinfo{year}{2022}), \bibinfo{pages}{37--50}.
\newblock


\bibitem[\protect\citeauthoryear{Zheng, Li, Zhang, Zhuang, Chen, Huang, Wang, Xu, Zhuo, Xing, Gonzalez, and Stoica}{Zheng et~al\mbox{.}}{2022a}]%
        {zheng2022alpa}
\bibfield{author}{\bibinfo{person}{Lianmin Zheng}, \bibinfo{person}{Zhuohan Li}, \bibinfo{person}{Hao Zhang}, \bibinfo{person}{Yonghao Zhuang}, \bibinfo{person}{Zhifeng Chen}, \bibinfo{person}{Yanping Huang}, \bibinfo{person}{Yida Wang}, \bibinfo{person}{Yuanzhong Xu}, \bibinfo{person}{Danyang Zhuo}, \bibinfo{person}{Eric~P. Xing}, \bibinfo{person}{Joseph~E. Gonzalez}, {and} \bibinfo{person}{Ion Stoica}.} \bibinfo{year}{2022}\natexlab{a}.
\newblock \showarticletitle{Alpa: Automating Inter- and Intra-Operator Parallelism for Distributed Deep Learning}. In \bibinfo{booktitle}{\emph{16th {USENIX} Symposium on Operating Systems Design and Implementation, {OSDI} 2022, Carlsbad, CA, USA, July 11-13, 2022}}, \bibfield{editor}{\bibinfo{person}{Marcos~K. Aguilera} {and} \bibinfo{person}{Hakim Weatherspoon}} (Eds.). \bibinfo{publisher}{{USENIX} Association}, \bibinfo{pages}{559--578}.
\newblock


\bibitem[\protect\citeauthoryear{Zheng}{Zheng}{2022}]%
        {SparTA-Github}
\bibfield{author}{\bibinfo{person}{Ningxin Zheng}.} \bibinfo{year}{2022}\natexlab{}.
\newblock \bibinfo{booktitle}{\emph{SparTA github}}.
\newblock
\urldef\tempurl%
\url{https://github.com/microsoft/SparTA/tree/sparta_artifact}
\showURL{%
\tempurl}


\bibitem[\protect\citeauthoryear{Zheng, Lin, Zhang, Ma, Yang, Yang, Wang, Yang, and Zhou}{Zheng et~al\mbox{.}}{2022b}]%
        {SparTA}
\bibfield{author}{\bibinfo{person}{Ningxin Zheng}, \bibinfo{person}{Bin Lin}, \bibinfo{person}{Quanlu Zhang}, \bibinfo{person}{Lingxiao Ma}, \bibinfo{person}{Yuqing Yang}, \bibinfo{person}{Fan Yang}, \bibinfo{person}{Yang Wang}, \bibinfo{person}{Mao Yang}, {and} \bibinfo{person}{Lidong Zhou}.} \bibinfo{year}{2022}\natexlab{b}.
\newblock \showarticletitle{$\{$SparTA$\}$:$\{$Deep-Learning$\}$ Model Sparsity via $\{$Tensor-with-Sparsity-Attribute$\}$}. In \bibinfo{booktitle}{\emph{16th USENIX Symposium on Operating Systems Design and Implementation (OSDI 22)}}. \bibinfo{pages}{213--232}.
\newblock


\end{thebibliography}

\end{document}